\documentclass[aps, onecolumn, showpacs, superscriptaddress,groupedaddress, nofootinbib, floatfix, 11pt]{revtex4-2}  

\usepackage{amsthm}
\usepackage{graphicx}  
\usepackage{dcolumn}   
\usepackage{bm}        
\usepackage{amssymb}
\usepackage{amsmath}   
\usepackage{hyperref}
\usepackage{float} 
\usepackage{color}
\allowdisplaybreaks

\hypersetup{colorlinks,linkcolor={blue},citecolor={blue},urlcolor={red}}
\def\beq{\begin{equation}}
\def\eeq{\end{equation}}
\def\barr{\begin{array}}
\def\earr{\end{array}}
\def\dis{\displaystyle}

\usepackage{appendix}
\DeclareMathOperator\arctanh{arctanh}

\def\d{{\rm d}}
\newcommand{\Mpl}{M_{\rm pl}}
\usepackage[normalem]{ulem}

\def\beq{\begin{equation}}
\def\eeq{\end{equation}}
\def\barr{\begin{array}}
\def\earr{\end{array}}
\def\dis{\displaystyle}

\definecolor{darkred}{cmyk}{0,1,1,0.4}

\begin{document}

\title{Universe bouncing its way to inflation}

\author{Manjeet Kaur}\email{
mkaur1@physics.du.ac.in}
\author{Debottam Nandi}\email{dnandi@physics.du.ac.in}
\author{ Debajyoti Choudhury}\email{debchou.physics@gmail.com}
\author{T. R. Seshadri}\email{trs@physics.du.ac.in}
\affiliation{Department of Physics and Astrophysics, University of Delhi, Delhi 110007, India}

\begin{abstract}
    Cosmological models with inflation and those with bounce have their own strengths and weaknesses. Here we construct a model in which a phase of bounce is followed by a viable inflationary phase. This incorporates several advantages of both and hence, is a more viable model for cosmic evolution. We explore scenarios wherein the bouncing phase smoothly transits to an inflationary one, with the pivot scale leaving the Hubble horizon during the latter era, thereby maintaining consistency with observations. Staying within the ambit of Einstein-Hilbert gravity augmented by the inflaton, we ensure a pre-inflationary bounce by introducing a second scalar field that helps engineer the requisite violation of the null energy condition. Potential ghost instabilities can be mitigated by invoking a non-trivial coupling between the two scalar fields.
    
    \vspace{4pt}
\end{abstract}

\maketitle
\section{Introduction}
The vanilla Big-Bang ansatz as a theory of the Universe, while consistent with general relativity and tying up with the observed Hubble expansion, has been found to be inadequate to address certain puzzles in the early Universe, such as the horizon problem, the flatness problem, to name a few. The inflationary paradigm, proposed as a brief period of accelerated expansion of the Universe in its early phase, not only overcomes these lacunae, but also explains known observations \cite{Guth:1981, LINDE1982389, Linde:1983gd,  STAROBINSKY198099, 1990-Kolb.Turner-Book, 2005hep.th....3203L, Sato:1981, Mukhanov:1981xt, HAWKING1982295, STAROBINSKY1982175, Guth:1982, VILENKIN1983527, Bardeen:1983, Albrecht-Steinhardt:1982, Mukhanov:1990me, Bassett:2005xm, Sriramkumar:2009kg, Baumann:2009ds, Linde:2014nna, Martin:2015dha, Albrecht:1982mp, Ade:2015lrj, Ade:2015ava}. An important example of the latter pertains to large-scale structure formation, which, within this paradigm, is supposed to have resulted from density perturbations arising from quantum fluctuations of the inflaton field. Observations (BICEP/Keck \cite{BICEP:2021xfz,Galloni:2022mok} and PLANCK \cite{Planck:2018jri}), suggest that, at the pivot scale ($k_*=0.05\, \text{Mpc}{}^{-1}$), the amplitude of the scalar power spectrum is $P_{s} \simeq 2.101^{+0.031}_{-0.034} \times 10^{-9} (68\%\, \, \text{CL})$ with the scalar spectral index being $n_s = 0.9649 \pm 0.0042\,(68\%\, \, \text{CL})$, while the ratio $(r)$ of the tensor-to-scalar perturbation amplitudes is bounded by $r<0.028\, (95 \%\, \, \text{CL})$. These observations agree remarkably well with the predictions of a near-scale-invariant spectrum of density perturbations in the inflationary paradigm. However, even with such successes, the inflationary paradigm leaves some questions unresolved, for example, ``the initial singularity problem,'' which necessarily pushes one into a regime where the semi-classical treatment (a standard staple of such analyses) itself is suspect \cite{Martin:2000xs}.

The classical bouncing paradigm \cite{PhysRevD.65.086007, Khoury:2001wf, Finelli:2001sr, Tsujikawa:2002qc, Brandenberger:2012zb, Brandenberger:1999sw, Novello:2008ra, Cai:2014bea, Battefeld:2014uga, Lilley:2015ksa, Ijjas:2015hcc, Brandenberger:2016vhg, Lehners:2008vx, Cai:2016hea, Brandenberger:2010bpq, Durrer:1995mz, Peter:2001fy}, as an alternative to inflation, seeks to efficiently solve the singularity problem, by postulating the Universe to have begun (albeit, in an unknowable state) in the asymptotic past with a non-zero scale factor, and undergoing a phase of contraction until the scale factor reaches a non-zero minimum, before entering an expanding phase that continues as of today. With the Hubble parameter ($H\equiv \dot a / a$, $a(t)$ being the scale factor of the Universe) changing its sign from negative to positive, a phase of $\dot{H}>0$ is imperative. On the other hand, in a flat Friedmann-Lema$\hat{i}$tre-Robertson-Walker (FLRW) Universe, we have $\dot{H}=-(\rho+p),$ where $\rho$ and $p$ are total energy and total pressure densities of the Universe, respectively. Thus, in such a Universe, a bounce needs $\rho + p <0$, which is a violation of the Null Energy Condition (NEC). Such a violation can be achieved, for example, using a scalar field with higher-order kinetic terms or with negative kinetic terms. This, typically leads to pathologies \cite{Rubakov:2014jja, Kobayashi:2016xpl, Libanov:2016kfc, Ijjas:2016vtq, Kolevatov:2017voe, Banerjee:2018svi, Cai:2016thi, Cai:2017dyi, Mironov:2018oec, Easson:2011zy, Sawicki:2012pz} such as ghost and gradient instabilities\footnote{It should be noted, however, that   single field non-singular bounce realizations exist within the framework of ghost-free generalized Galileon cosmology \cite{Qiu:2013eoa, Qiu:2011cy, Osipov:2013ssa}, provided higher-derivative terms are included. Such models typically  require the introduction of a  second field in order to achieve near-scale-invariant  power spectrum thereby satisfying current observational constraints. Rather than adopting this route of higher-derivative terms, in this article, we restrict ourselves to two scalar fields (coupling minimally to gravity) and with the simplest kinetic term. Further, we demonstrate a detailed post-bounce scenario in order to obtain a near-scale-invariant spectrum. As we move forward, the corresponding details will become increasingly evident.}. 

The classical bounce epoch is, of course, characterized by a Hubble parameter $H$ that vanishes at the bounce epoch. While different functional forms of $H$ may be postulated (or even derived if one were to postulate, instead, a fundamental theory), it is instructive to view the Hubble parameter in terms of two, as yet unknown, functions
$f_{1,2}(a)$,  {\em viz.}
\begin{eqnarray}\label{Eq:genbounceH}
	H^2=f_1(a)-f_2(a)\ .
\end{eqnarray}
The functions $f_{1,2}(a)$ are, by definition, positive semi-definite, and $f_1\left(a\right) \geq f_2\left(a\right)$ with the epoch of equality characterizing the bounce. The superficial resemblance of Eq. \eqref{Eq:genbounceH} with the Friedmann equation suggests that a two-field theory might profitably be used in achieving this, with $f_{1}(a)$ and $-f_{2}(a)$ being identified with the respective energy densities associated with the two fields. 
To minimize potentially deleterious consequences, $f_2(a)$ may be chosen such that, during the entire evolution of the Universe, it contributes significantly only close to the bounce.
A toy example is suggested by power-law evolution of the scale factor ({\em viz.}, $a(t) \sim t^\alpha$), ubiquitous in cases where the energy density is dominated by a single perfect fluid. This naturally leads to $H \propto a^{-1/\alpha}$. In the spirit of Eq. \eqref{Eq:genbounceH}, let us consider, then
\begin{eqnarray}\label{eq:hubblematterbounceeg}
	H^2\sim \left(\frac{a_0}{a}\right)^m-\left(\frac{a_0}{a}\right)^n\, ,\qquad \, n>m \ .
\end{eqnarray}
Here, $a_0$ is the value of scale factor at bounce. For $a \gg a_0$, the first term on the right-hand side dominates, leading to the power law evolution of the scale factor $(a\propto t^{2/m})$. Assuming an initial contracting phase, as the scale factor approaches $ a_0$, the second term on the right-hand side becomes relevant, leading to a smooth transition from contraction to expansion. Further specializing to $m = 3$, commonly referred to as the matter bounce scenario, at late times, an adiabatic perturbation would give rise to an exact scale-invariant scalar power spectrum  \cite{Brandenberger:2012zb, 1970ApJ...162..815P, 1970Ap&SS...7....3S} rendering it phenomenologically interesting\footnote{However, while observations suggest near scale-invariant scalar perturbations, the measured scalar spectral index is decidedly different from unity.}. However, a fine-tuning of initial conditions is often necessary in such a model. Exceptions are attractors, such as ekpyrotic models \cite{Garfinkle:2008ei, Levy:2015awa} with $m > 6$. These, however, are not in line with observations \cite{Planck:2018jri, Aghanim:2018eyx}. 

A relatively straightforward way out of the impasse would be to move away from the simplistic scenario of Eq. \eqref{eq:hubblematterbounceeg} and consider, instead, a form of $f_1(a)$, such that well after the bounce, $f_1(a)$ has a form that admits a rapid expansion. An example is afforded by the chaotic inflationary scenario wherein, during the slow-roll epoch, the Hubble expansion approximates to 
\begin{eqnarray}\label{eq:chaotic-intro}
	H^2\propto \left(C_1 +C_2 \ln{\left(\frac{a}{a_i}\right)}\right),\ C_1>0,\ C_2<0,\ a_i>0,
\end{eqnarray}
where $C_{1,2}$ and $a_i$ are appropriate constants (for details, see \ref{appendix_B}).
Note that, notwithstanding the example of Eq. \eqref{eq:hubblematterbounceeg}, it is not imperative that $f_2(a)$ should have a form similar to that for $f_1(a)$. Indeed, if we ascribe $f_{1,2}(a)$ to different fields, their functional forms would be determined by the respective field Lagrangians. Thus, ascribing $f_1(a)$ to the form above while maintaining the power law form for $f_2(a)$ could still lead to a bounce for appropriate choices of the parameters. Such a choice has the distinct advantage that post-bounce, $f_2(a)$ falls off quickly in comparison to $f_1(a)$, and the latter leads to a phase of post-bounce inflation. This would not only solve issues such as the horizon problem but also provide an observationally viable seed for large-scale structure formation.

A further advantage of the existence of an inflationary phase is that it serves to dilute possible effects (such as the generation of inherently anisotropic perturbations) of the second field that is necessary to drive the bounce. Together, we would, then, have all the good features of bounce and inflation (and problems of neither), with one phase going over smoothly to the other.

One approach to achieving such evolution involves modifying the laws of gravity at high energy scales, for example, theories like Loop Quantum Cosmology \cite{Barboza:2020jux, Gegenberg:2016vhx}, modifications to general relativity \cite{Bamba:2016gbu, Saidov:2010wx, Daniel:2022ppp}, and scalar fields coupled non minimally to gravity \cite{Qiu:2014nla, Xia:2014tda, Ni:2017jxw, Qiu:2015nha, Wan:2015hya, Sloan:2019jyl, Mathew:2018rzn, Upadhyay:2023wcq}.
Rather than adopting this route, our goal is to construct a two-scalar-field scenario (with each coupling minimally to gravity) that naturally leads to the aforementioned features. While we should establish that this is indeed possible without resorting to undue fine-tuning, the discerning reader would point out that a ghost instability --- a consequence of the violation of the NEC --- may still plague us. Admittedly, this cannot be entirely avoided in a construction such as ours. However, the said violation can be confined to a very small time window (around the bounce) if one would consider an appropriate form of $f_2(a)$ that is positive only in the vicinity of the bounce and negative otherwise. We would also demonstrate that this can come about naturally \emph{e.g.}, by introducing an appropriate coupling function in a large class of two-field models. As for the violation remaining close to the bounce yet, we have no answer at present. It is possible that the final resolution lies in quantum corrections.

The structure of this paper is as follows. In Sec. \ref{sec:1-General-Formalism}, we provide the general equations required to construct a bouncing solution. In Sec. \ref{sec:3-inflation}, we focus on constructing a bouncing phase followed by the inflationary dynamics. We do this in two steps. To begin with, we demonstrate how two popular inflationary models can be tweaked (albeit in an ad-hoc manner) to achieve this goal. Thereafter, we show how a simple augmentation of a very well-motivated model also leads to very similar results.
As these models suffer from a ghost instability, we propose, in Sec. \ref{sec:4-removing-ghost}, a way to mitigate the problem by invoking a coupling between the two fields. Finally, in Sec. \ref{sec:6-conclusion}, we summarise and present a future outlook.
We work with the natural units of  $\hbar=c=1$ and define the Planck mass to be $\Mpl \equiv (8\pi G)^{-1/2}= 1$ (In other words, we rescale dynamical fields and parameters of mass-dimension +1 in terms of the Planck mass $\Mpl$). Adopting the metric signature $(-,+,+,+)$, we use the e-N-folds variable $\mathcal{N}$, defined as
\beq
\mathcal{N}\equiv\pm \sqrt{2\ln{(a/a_0)}},
\eeq
interchangeably for cosmic time $t$, with the negative (positive) branches referring to the contracting (expanding) phases.
Overdot $( \dot{\, \, })$  and overprime $(\,'\,)$ denote derivatives with respect to cosmic time $t$, and $\mathcal{N}$, respectively. 
\section{Preliminaries }\label{sec:1-General-Formalism}
We begin by setting up the generic formulation, with details postponed to subsequent sections. The large-scale spatial homogeneity and isotropy of the Universe prompts the use of the flat FLRW line element, namely
\begin{equation}\label{Eq:FLRWmetric}
	\d s^2 
	= -\d t^2+a^2(t)\, \d \textbf{x} ^2.
\end{equation}

\noindent The goal being to generate a form of $a(t)$ consistent with a bouncing Universe, the matter content must be chosen appropriately.  However, with a bounce seemingly requiring a violation of the NEC, this cannot be achieved with a single scalar field that is minimally coupled to gravity. While it is possible that modifying the usual Einstein-Hilbert action for gravity may lead to a bounce, we eschew it on account of various complications (instabilities) that it often entails and restrict ourselves to minimally coupled scalar fields along with the canonical gravity action. As can be expected, a form such as Eq. \eqref{Eq:genbounceH} is most easily achieved if we associate $f_{1,2}(a)$ with two different scalar fields, each with commensurate action. Starting with the simplest theory, we posit an
action
\begin{eqnarray}\label{eq:actiongenP1P2}
	&\mathcal{S}=\frac{\Mpl^2}{2}\int{\rm d}^4{\rm \bf x} \sqrt{-g} \left(R-2L_1(\phi,\partial_\mu \phi)-2L_2(\chi,\partial_\mu \chi)\right),&\nonumber\\
	&&\hspace{-1mm}
\end{eqnarray}
with $L_1(\phi,\partial_\mu \phi)$ and $L_2(\chi,\partial_\mu \chi)$ being the respective Lagrangian densities\footnote{Here, fields $\phi$ and $\chi$ are dimensionless, and Lagrangian densities are of mass dimension $+2$.}. The separability of the Lagrangian\footnote{While this assumption might seem strange in the absence of symmetry, it is technically natural as long as quantum corrections accruing from their putative interactions with other fields can be neglected. However, this is only a simplification, and, at a later stage, we would actually relax this restriction} allows us to associate $\phi \leftrightarrow f_1(a)$ and $\chi \leftrightarrow f_2(a)$ with the dynamics of the Universe being dominated by the respective fields in different epochs. While the Lagrangian for $\phi$ would be the canonical one, that for $\chi$ must have the wrong sign (negative) for the kinetic term, {\em viz.},
\begin{eqnarray}
	\label{eq:canonicalP1}
	L_1(\phi,\partial_\mu \phi) &=& \ \ \frac{1}{2}\partial_\mu \phi \partial^\mu \phi + V_1(\phi),\\    L_2(\chi,\partial_\mu \chi)&=&-\frac{1}{2}\partial_\mu \chi \partial^\mu \chi + V_2(\chi).
\end{eqnarray}
Such a choice for $L_2$ (also used in \cite{Cai:2007qw}) would obviously lead to ghost instabilities. However, for the sake of simplicity, we neglect this for now and would return to it at a later point.

For the action in Eq. \eqref{eq:actiongenP1P2}, the equations of motion are
\begin{eqnarray}
	G^\mu_\nu=T^\mu_{\nu(\phi)}+T^\mu_{\nu(\chi)},\quad\nabla_\mu T^{\mu\nu}_{(\phi)}=0,\quad\nabla_\mu T^{\mu\nu}_{(\chi)}=0,
\end{eqnarray}
where $T^\mu_{\nu(\phi)}$ and $T^\mu_{\nu(\chi)}$ are the
stress-energy tensors corresponding to the two fields, and leads to
\begin{eqnarray}\label{eq:genhubfortwofields}
	3 H^2 = \rho_\phi +\rho_\chi, \qquad
	-2\dot{H} = \rho_\phi+P_\phi+\rho_\chi+P_\chi,
\end{eqnarray}
where the respective energy  and pressure densities are given by
\begin{eqnarray}\label{eq:rhophi_pphi}
	\rho_\phi&=& \frac{1}{2}\dot{\phi}^2 +V_1(\phi),\quad\qquad P_\phi= \frac{1}{2}\dot{\phi}^2 - V_1(\phi),\\\label{eq:rhochi_pchi}
	\rho_\chi&=& -\frac{1}{2}\dot{\chi}^2 + V_2(\chi),\quad\quad  P_\chi= -\frac{1}{2} \dot{\chi}^2 - V_2(\chi).
\end{eqnarray}
In terms of the e-N-fold variable $\mathcal{N}$, Eq. \eqref{eq:genhubfortwofields} can be rewritten as
\begin{eqnarray}\label{eq:eqn1_in_e_N_fold}
	H^2 = 2\mathcal{N}^2 \frac{V_1(\phi) + V_2(\chi)}{6\mathcal{N}^2-\phi^{\prime 2}+\chi^{\prime 2}},\quad\quad     \frac{H'}{H}=\frac{\chi^{\prime 2}-\phi^{\prime 2}}{2\mathcal{N}}.
\end{eqnarray}
The equations of motion for the scalar fields \emph{viz}
\begin{eqnarray}
	\ddot{\phi} + 3H \dot{\phi}+V_{1,\phi}=0,\qquad \quad \ddot{\chi} + 3H \dot{\chi} - V_{2,\chi}=0,
\end{eqnarray}
(where $A_{,x}\equiv \partial A/\partial x$), can, similarly,  be reexpressed as
\begin{eqnarray}\label{eq:eq_of_motion_phi_chi}
	\phi'' + \left( 3 \mathcal{N} + \frac{ H'}{H}-\frac{1}{\mathcal{N}}\right)\phi'+\frac{\mathcal{N}^2}{H^2}V_{1,\phi}&=&0,\\ \chi'' + \left( 3 \mathcal{N} + \frac{ H'}{H}-\frac{1}{\mathcal{N}}\right)\chi' - \frac{\mathcal{N}^2}{H^2}V_{2,\chi}&=&0.
\end{eqnarray}
\section{Pre-inflationary bounce}\label{sec:3-inflation}

Since we want the bounce to be immediately followed by an inflationary epoch, the most optimal choice for $V_1(\phi)$ would be one that admits the requisite slow-roll. The simplifying assumption --- of the dynamics of the $\chi$ and $\phi$ fields being largely decoupled allows us not only to choose $V_1(\phi)$ almost independent of $\chi$ but also to choose a $V_2(\chi)$ that facilitates a bounce. Such a separation of dynamics, though, is not expected to hold as the Universe enters into or exits from a phase where the contribution of the $\chi$ field is comparable to (albeit less than) that of the $\phi$ field. The dynamics during these epochs is not analytically tractable, and one needs to take recourse to numerical methods. Nevertheless, to gain insight, we begin by splitting the domain of interest into three different parts: the contraction, the bounce, and the expansion. To qualitatively understand the features of the dynamics, we begin by considering a representative scenario, namely chaotic inflation \cite{Linde:1983gd}. 
Any model must not only yield bounce as well as sufficient slow-roll inflation, but also must be consistent
with the plethora of accurate cosmological observations. 
The classical chaotic inflation scenario, of course, does not admit a bounce, and to introduce it the potential would need to be altered. 
\subsection{Bounce with Chaotic inflation}\label{sec:3a-chaotic-inflation}
A real scalar field $\phi$ with a canonical kinetic term, minimally coupled to gravity, and with the simple quadratic potential
\begin{eqnarray}{\label{eq:potential_chaotic}}
	V(\phi)=\frac{1}{2}m^2 \phi^2,
\end{eqnarray}
is well known to lead to a phase of chaotic inflation, provided $m \ll 1$ (necessary for an observationally acceptable amplitude of the power spectrum) and one starts with a large enough value for the field, {\em viz.} $|\phi| \gg 1$ (necessary for an adequate duration of slow-roll). Furthermore, with $V(\phi) \ll 1 $ automatically ensuring that the energy content is smaller than the Planck mass (see, for example, \ref{appendix_D}), the model still admits a semi-classical treatment and a trans-Planckian phase is avoided. The corresponding solutions for the scalar field as well as the Hubble parameter, as applicable well within the slow-roll regime (see, for details, \ref{appendix_D}), can be approximated as
\begin{eqnarray} \label{eq:CIphi}
	\barr{rcl}
	\phi^2\simeq\dis \xi_i -2\mathcal{N}^2 \ ,\quad
	3H_{(\phi)}^2 \simeq \dis V(\phi) \simeq \frac{m^2}{2}(\xi_i -2\mathcal{N}^2),
	\earr
\end{eqnarray}
where, $\phi_i \equiv \phi(\mathcal{N}_i)$ describes the initial condition, $\xi_i \equiv \dis 2\mathcal{N}_i^2+\phi_i^2$, and $H_{(\phi)}$ denotes that only the single scalar $\phi$ is included in the description. Since the bulk of the expansion presumably occurs during the inflationary era, it is reasonable to identify the epoch of bounce with $\mathcal{N}=0$. Now, with the slow-roll parameter being given by $\epsilon\equiv-\dot{H}/H^2=({\rm d}\phi/\rm {d} \mathcal{N})^2/(2\mathcal{N}^2)$, the end of inflation is signalled by ${\rm d}\phi/\rm {d} \mathcal{N}\approx \pm \sqrt{2}\mathcal{N}$. Used along with ${\rm d}\phi/{\rm {d}} \mathcal{N} = - \mathcal{N}({\rm d}\ln{V} /{\rm d}\phi$), this translates to $\phi_{\rm end}\approx \pm\sqrt{2}$. The first of the Eqs. \eqref{eq:CIphi}, then, implies that, for sufficient inflation to have taken place, $\mathcal{N}_{\rm end}\approx\phi_i/\sqrt{2}$. In other words, one must have $\xi_i\gg2$ for phenomenological acceptability; for $\xi_i<2$, there is too little of the slow roll, and the field quickly moves to a phase of rapid oscillations \cite{2005-05-14-MukhanovViatcheslav-Physicalfoundationsofcosmology}.

On inclusion of a second scalar field $\chi$, especially with a non-canonical kinetic term,  Eq. \eqref{eq:CIphi} would, of course, change. Considering, for illustrative purposes, the form of Eq. \eqref{Eq:genbounceH}, we would, now, have
\begin{eqnarray}\label{eq:infbouncehub}
	3H^2 = 3 H_{(\phi)}^2 - \left(\frac{a_0}{a}\right)^n
	=   3 H_{(\phi)}^2 - \frac{m^2}{2} \xi_i e^{-n \mathcal{N}^2/2} \ ,
\end{eqnarray}
where, once again, $H(\mathcal{N} = 0) = 0.$ The quest, then, devolves to finding potentials that lead to such solutions. As mentioned earlier, we take recourse to splitting the problem into three-time domains: the contracting, bouncing, and expanding phases. If the potential happens to be symmetric about $\phi=\phi_{bounce}$,  an expanding solution can be mapped back into a contracting one by simply turning the clock backward (see Eq. \eqref{eq:infbouncehub}), and, hence, we return to the contracting phase later, and begin with the bounce.

\subsubsection{Bouncing phase}

The (self-imposed) requirement of Eq. \eqref{eq:infbouncehub} being satisfied does not exhaust the possibilities for the potentials $V_1(\phi)$ and $V_2(\chi)$. However, further assuming that the dynamics of the Universe is entirely separable in terms of the $\phi$-driven and the $\chi$-driven, allows us to narrow down on the time-variation of the fields as well as the potentials (implicitly through the fields)\footnote{Note that these are only sufficient conditions, and not necessary ones.}, namely

\begin{eqnarray}
	\phi' &= & \dis \frac{ -2\mathcal{N}}
	{\left(-2\mathcal{N}^2+ \xi_i \, \left(1 - e^{-n\mathcal{N}^2/2}\right)\right)^{1/2}},
	\\
	V_1(\mathcal{N}) & = & \dis
	\frac{m^2}{6}\left(-2-6\mathcal{N}^2+3 \xi_i\right),
	\\
	\chi' & = & \dis
	\frac{\sqrt{n~ \xi_i} \; e^{- n \mathcal{N}^2/4}}
	{\left(-2\mathcal{N}^2+ \xi_i \, \left(1 - e^{-n\mathcal{N}^2/2}\right)\right)^{1/2}},\\
	V_2(\mathcal{N}) & = & \dis \frac{m^2}{12}\, (n-6)\, \xi_i \,
	e^{-n\mathcal{N}^2/2} \ .
\end{eqnarray}
Here, $V_1(\mathcal{N})\equiv V_1(\phi(\mathcal{N}))$ and $V_2(\mathcal{N})\equiv V_2(\chi(\mathcal{N}))$.
Integrating the equations for $\phi$ and $\chi$ does not yield closed-form analytical solutions. However, for $B_c  \equiv  \dis n \xi_i/4-1>0$ (easily satisfied for $\xi_i\gg2$, as we must have), the small-$\mathcal{N}$ solutions can be approximated as
\begin{eqnarray}
	\phi(\mathcal{N}) &\simeq& \phi_0 -  \, \mathcal{N} \; \sqrt{\frac{2 }{B_c}} \ ,\\
	\chi(\mathcal{N}) &\simeq& \chi_0 +  \, \mathcal{N}
	\sqrt{\frac{3 \, \xi_i}{B_c}},
\end{eqnarray}
Within the regime of validity of this approximation (i.e., the period wherein $\mathcal{N}<1$), the potentials can be approximated by
\begin{eqnarray}
	\label{eq:pot_phi_chaotic}
	V_1(\phi) &\approx& \dis \frac{m^2}{2} \left(A_c-B_c(\phi-\phi_0)^2\right)
	,\\
	V_2(\chi) &\approx& \dis \frac{m^2}{12} (n - 6) \xi_i \,
	\exp\left(-\frac{n B_c}{6\xi_i} \, (\chi -\chi_0)^2 \right),\ 
\end{eqnarray}
where $A_c \equiv \dis  \xi_i - 2/3 $. Although a nonzero value of $A_c$ seems to indicate a cosmological constant, in contrast to Eq. \eqref{eq:potential_chaotic}, it needs to be realized that the form of $V_1(\phi)$ in Eq. \eqref{eq:pot_phi_chaotic} is only a local expansion and is not valid over the entire range of $\phi$. The potential for $\chi$ has a Gaussian profile; for  $n \xi_i \gg 4$, the width scales as $n^{- 1}$ (since, in this limit, $B_c\simeq n\xi_i/4$) and increasing $n$ serves to reduce the duration during which $V_2(\chi)$ plays a significant role. Interestingly, even a vanishing potential ($n = 6$) is admissible.

\subsubsection{Expanding phase}
In the current illustrative example, designed so that the contribution of the field $\chi$ can be disregarded well outside of the bouncing regime, the dynamics therein is almost completely dominated by the inflaton field $\phi$. Hence, the results for chaotic inflation, namely Eq. \eqref{eq:CIphi}, would follow provided the potential for the scalar field $\phi$, as applicable well within this regime, can be approximated by Eq. \eqref{eq:potential_chaotic}.
\noindent Defining $\phi = \phi_i$ ($|\phi_0|>|\phi_i| \gg 1$) to signal the onset of the slow-roll inflation, as $|\phi|$ decreases below $|\phi_i|$, the Universe inflates ($\mathcal{N}$ increases from $\mathcal{N}_i$) and, finally at $|\phi|\simeq\sqrt{2}$, inflation ends. For $|\phi| < \sqrt{2}$, the field begins to oscillate around the bottom of the potential, resulting in the classic preheating scenario.  

With $\phi = \phi_i$ denoting the boundary between the bounce and the inflationary phase --- and the two local parameterizations of the potential $V_1(\phi)$ --- matching conditions (continuity and smoothness) must be imposed. This leads to

\begin{eqnarray}\label{eq:phii}
	\phi_i&=&\frac{\sqrt{n}(-2 + 3\phi_0^2)+\zeta}{6 \sqrt{n}\, \phi_0},\\ \mathcal{N}_i^2&=&\frac{36\phi_0^2+n(-2+3\phi_0^2)+\sqrt{n}\zeta}{18 n \phi_0^2}.
\end{eqnarray}

where $\zeta=\sqrt{-144 \phi_0^2 +n(2+3\phi_0^2)^2}$, and both $\phi_i$ and $\mathcal{N}_i$ are real as long as
\begin{eqnarray}
	\left(\phi_0-\frac{2}{\sqrt{n}}\right)^2\geq \frac{4}{n}-\frac{2}{3}.
\end{eqnarray}
Further, for $\phi_0 \gg 1$ (as required to achieve slow-roll inflation),
$$\phi_i \rightarrow \phi_0, \quad \mathcal{N}_i^2 \rightarrow \frac{1}{3} + \frac{2}{n}.$$ 
As long as $n>3$, it is guaranteed that the inflationary phase starts with $\mathcal{N}_i<1$. 
\subsubsection{Contracting phase}
\begin{figure}[t!]
	\includegraphics[height=0.42\textwidth,width=0.495\textwidth]{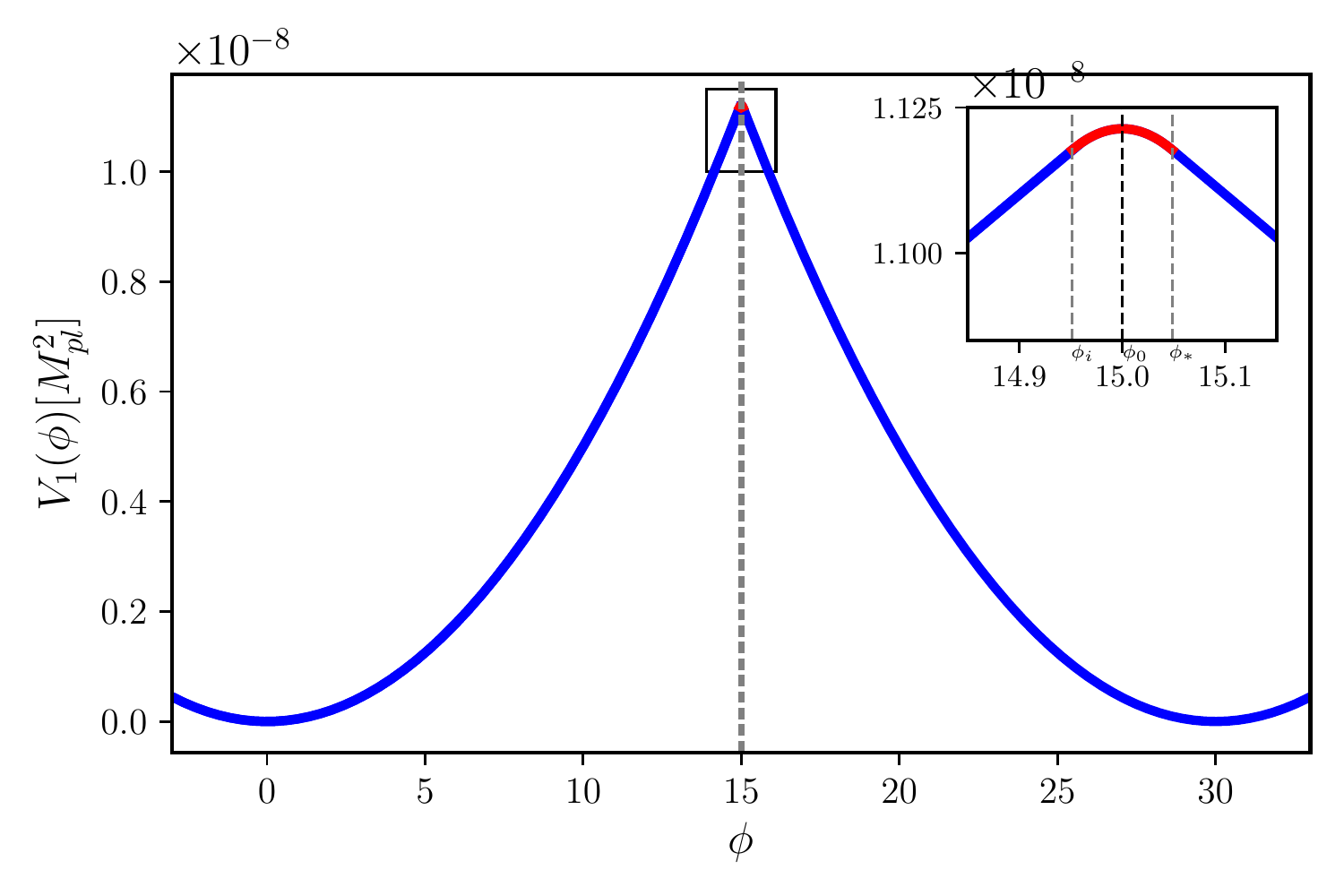}
	\includegraphics[height=0.4\textwidth,width=0.495\textwidth]{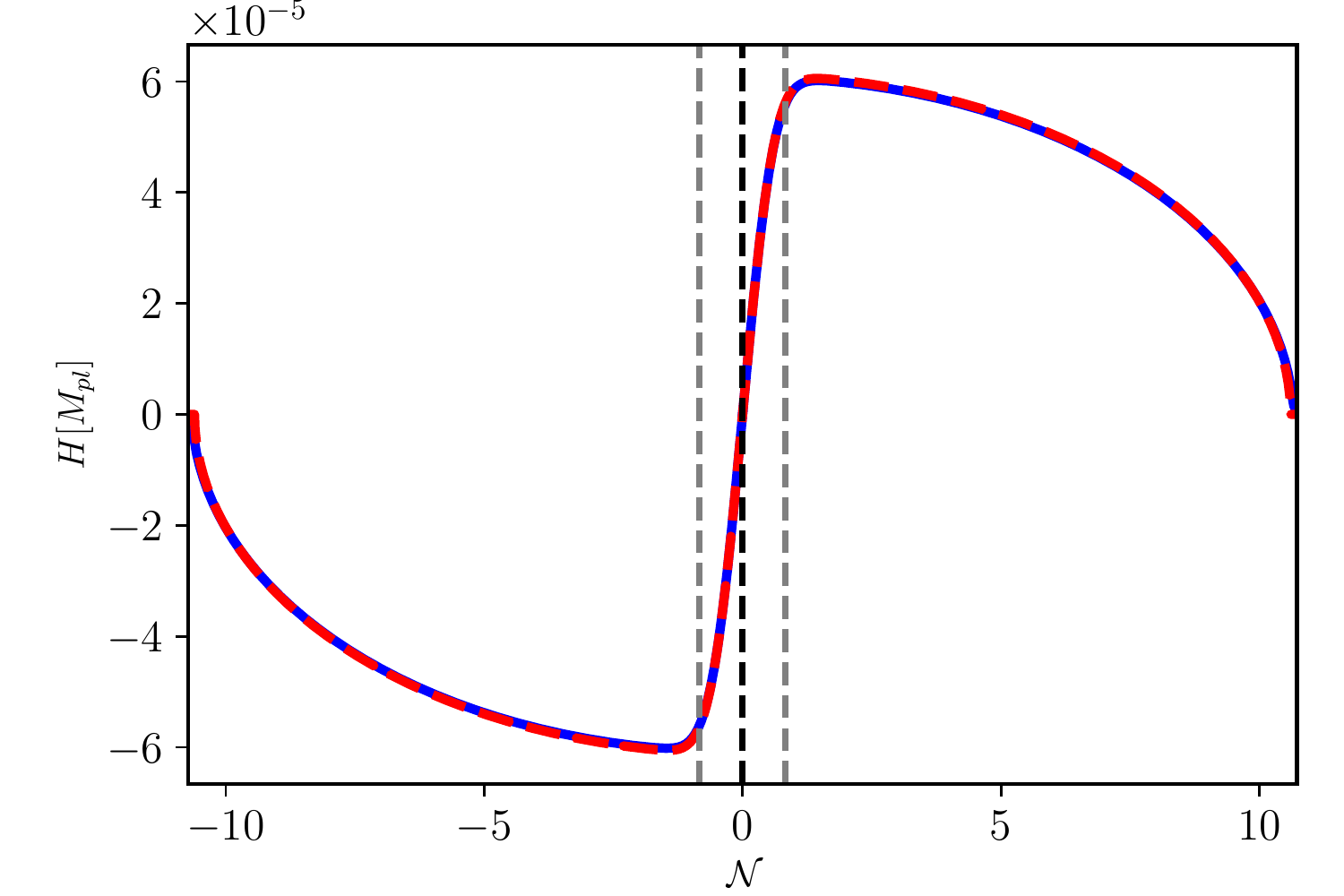}
	\caption{(Left) the potential $V_1(\phi)$ described in Eq. \eqref{eq:pot_chaotic_inf_bounce} for the parameters, $m=10^{-5}$, $n=5.5$, $\phi_0=15$. The red part in the inset corresponds to the bouncing phase, while the blue parts to the right (left) correspond to the contraction (expansion) phases, respectively. (Right) the corresponding evolution of the Hubble parameter $H$  with $\mathcal{N}$, as appropriate for $\chi_0=0$. The solid (blue) line corresponds to the approximate analytical solution given by Eq. \eqref{eq:infbouncehubconraction}, and the red (dashed) line corresponds to the numerical solution. As can be seen, the approximate analytical solution is almost indistinguishable from the full numerical solution. The black (dashed) line corresponds to the epoch of bounce with $\mathcal{N}=0$, and the left (right) gray (dashed) lines correspond to the transition from the contracting to the bounce (bounce to inflating) phases with $|\mathcal{N}|=0.85$. }
\label{fig:potential_chaotic_inflation}
\end{figure}

Since the history of the contracting phase is essentially wiped out by inflation, one could, in principle, choose any of a very wide class of potentials to drive the corresponding evolution. It is instructive to demand that the potential be symmetric around the bounce, thereby leading to a symmetry between the contracting and the expanding phase\footnote{We would see later that this restriction may lead to phenomenological inconsistencies and would be relaxed.}. Defining $\tilde{\phi}\equiv \phi-\phi_0$, the potential, over the entire range of $\tilde\phi$ can, then, be written as
\begin{eqnarray}\label{eq:pot_chaotic_inf_bounce}
	    &&V_1(\phi) =
		\left\{
		\barr{lclcl}
		\dis \frac{1}{2}m^2 (\tilde{\phi}-\phi_0)^2, &\qquad\quad&\tilde{\phi} \geq \phi_0-\phi_i, &\qquad\quad& (\text{Slow-roll contraction})
		\\[1.5ex]
		\dis   A_c - B_c\tilde{\phi}^2, & & |\tilde{\phi}| < \phi_0 -\phi_i, && (\text{Bounce})
		\\[1.5ex]
		\dis \frac{1}{2}m^2 (\tilde{\phi}+\phi_0)^2, && \tilde{\phi} \leq \phi_i-\phi_0,  && (\text{Slow-roll expansion}),
		\earr
		\right.\\
		&&\text{and}\qquad\qquad\qquad\nonumber\\
		&&V_2(\chi) = \dis \frac{m^2}{12} (n - 6) \xi_i \,
		\exp\left(-\frac{n B_c}{6\xi_i} \, (\chi -\chi_0)^2 \right),
\end{eqnarray}

\noindent as is depicted in Fig. \ref{fig:potential_chaotic_inflation} for a particular choice of parameters. Up to the end of Inflation $(\mathcal{N}<\phi_i/\sqrt{2})$, the Hubble parameter takes the form

\begin{eqnarray}\label{eq:infbouncehubconraction}
	3H^2 \equiv \begin{cases}
		\frac{1}{2}m^2(-2\mathcal{N}^2+\xi_0)-\frac{1}{2}m^2\xi_0e^{\frac{-n \mathcal{N}^2}{2}},\quad &\mathcal{N} \leq 0, \\
		\frac{1}{2}m^2(-2\mathcal{N}^2+\xi_i)-\frac{1}{2}m^2\xi_ie^{\frac{-n \mathcal{N}^2}{2}}, \quad & \mathcal{N}\geq 0,
	\end{cases}, \quad \xi_0 \equiv \dis 2\mathcal{N}_i^2+(2\phi_0-\phi_i)^2.\nonumber\\
\end{eqnarray}

and is depicted in Fig. \ref{fig:potential_chaotic_inflation}. Here, for this choice, the bounce period lies between $-0.85\leq\mathcal{N}\leq0.85$, followed by inflation till $ \mathcal{N}\sim 11$.

The parameters of the theory are to be chosen so that the slow roll dynamics dominates the inflationary epoch, at least until when the pivot scale ($0.05\, \text{Mpc}^{-1}$) leaves the Hubble horizon. This helps ensure that the scalar and the tensor perturbations behave similarly to those in standard slow-roll inflation.
In other words, $\mathcal{N}_{end}^2-\mathcal{N}_*^2$ (where $\mathcal{N}_{end}$ and $\mathcal{N}_*$ are the e-N-folding numbers corresponding to the end of inflation and the pivot scale respectively),  must be\footnote{The constraint on duration of at least 50 e-folds of inflation suggests  $\mathcal{N}_{end}^2-\mathcal{N}_*^2 \geq 100$.} $\sim 100$. In the case of chaotic inflation, the above condition corresponds to $\phi_0\geq 15$, whereas in the case of Starobinsky inflation (which will be discussed in detail in the next section), it corresponds to $\phi_0\geq 6$.

It is, thus, quite apparent that such a model would not only produce a bounce but also lead to sufficient inflation. However, the very structure of the potential given by Eq. \eqref{eq:pot_chaotic_inf_bounce} seems to be an ad-hoc one. Furthermore, observations also suggest that small-field models are preferred over large-field ones. We next address both these issues within a popular small-field scenario.
\subsection{Bounce with Natural inflation}\label{sec:Natural_bounce}
In the toy model that we studied above, the bounce phase coincides with the epoch wherein $V_1(\phi)$ is convex upwards. This is not unexpected, for rolling down a potential causes expansion while going up a potential hill typically results in contraction (or, at the very least, a slowing down of a pre-existing expansion). Thus, the transition from a contracting phase to an expanding one, while primarily driven by a different field $\chi$, should be accompanied by a $V_1(\phi)$ that is (at least locally) convex upwards. Of course, for the inflation to be arrested (as well as for a stable vacuum to exist), this downward spiraling of $V_1(\phi)$ needs to be curtailed, which led us to the patch-wise definition of the potentials. However, a potential, fulfilling such criterion already exists in the literature \cite{PhysRevLett.65.3233}.

\begin{figure}[t!]
	\centering
	\includegraphics[height=.45\textwidth,width=.6\textwidth]{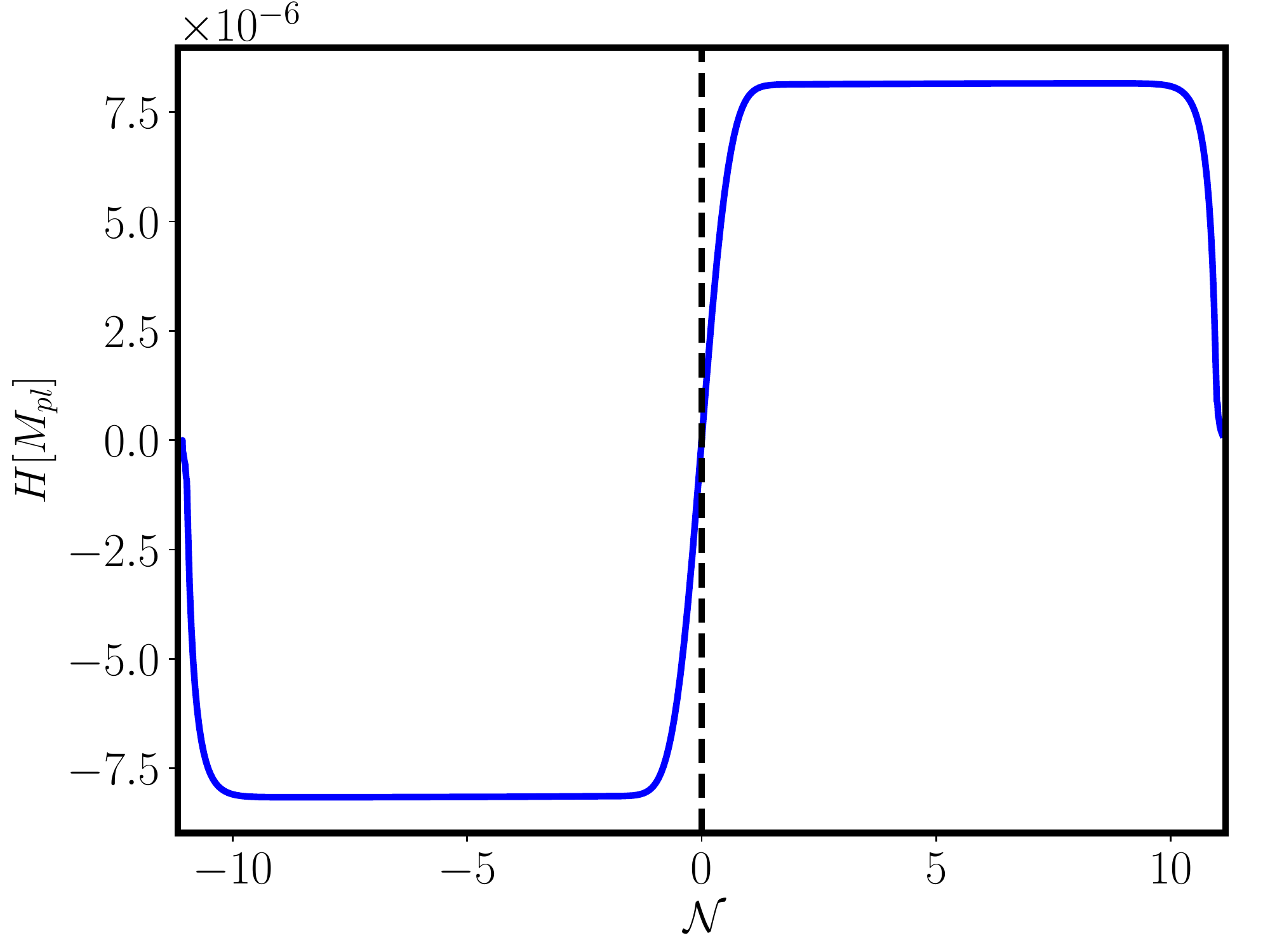} 
	\caption{The evolution of the Hubble parameter $H$  with $\mathcal{N}$, obtained for the potentials described by Eqs. \eqref{eq:natural_inf_pot_phi} and \eqref{eq:natural_inf_pot_chi} and parameter values $\lambda=10^{-10}$, $n=5.5$ and $\mu=1.5 $. The dashed line corresponds to the epoch of bounce.}
	\label{fig:hubble_num_natural_inflation}
\end{figure}

The twin requirements of generating both an adequate amount of expansion as well as consistent primordial perturbation necessitate that the ratio of the potential’s height to its width must satisfy
\begin{eqnarray}
	\frac{\Delta V}{(\Delta\phi)^4}\leq \mathcal{O}\left(10^{-6}-10^{-8}\right),
\end{eqnarray}
where $\Delta V$ is the change in the inflationary potential and $\Delta \phi$ is the change in the inflaton field $\phi$ during the slow-roll. This extreme ratio between mass scales referred to as a ``fine-tuning” problem \cite{Bassett:2005xm,kinney2009tasi, Baumann:2009ds} is ameliorated in Natural Inflation \cite{PhysRevLett.65.3233}, wherein the flatness arises naturally on account of a shift symmetry being imposed. A breaking of the symmetry, at a scale $\mu$, results in a small potential for this pseudo-Nambu-Goldstone boson of the form
\begin{eqnarray} \label{eq:natural_inf_pot_phi}
	V_1(\phi)=\lambda\left(1-\cos{\left(\frac{\phi}{\mu}\right)}\right) \ ,
\end{eqnarray}
where the height is naturally much smaller than the width, i.e., $2\lambda \ll \pi \mu$. Adopting this scenario, we augment it by the aforementioned $\chi$ field with the wrong sign for the kinetic energy, once again ascribing it to the potential
\begin{eqnarray}
	\label{eq:natural_inf_pot_chi}
	V_2(\chi) &=& P_n \exp\left(-Q_n\left(\chi -\chi_0\right)\right)^2,
\end{eqnarray}
where $P_n = -\lambda\left(\frac{n-6}{3}\right)$ and $Q_n=\frac{n}{4}.$
The resultant bouncing solution for $H(\mathcal{N})$ is shown in Fig. \ref{fig:hubble_num_natural_inflation}.
As in the preceding case (the chaotic-inflation-inspired scenario), here too, both bounce and sufficient inflation are obtained. This scenario has the added bonus that, instead of the patchwork potential of Eq. \eqref{eq:pot_chaotic_inf_bounce}, the potential here appears naturally in a model with a shift symmetry and meets all the requirements. Nominally, then, this could be a good solution to the problem at hand. However, certain important issues remain, and this has to do with the agreement with the observational data. The Cosine Natural Inflation model is consistent with the recent Planck observational data as long as $\mu\geq 1.5$. As for Planck (TT, TE, and EE correlation as well as Low E) and Lensing data \cite{Planck:2018jri, Aghanim:2018eyx}, the agreement is only at $2\sigma$ level. However, it is inconsistent with the BAO + BK18 observations \cite{ BICEP:2021xfz}. It has been argued, that it is possible to reconcile such differences by modifying the original theory (such as introducing additional scalar fields or non-minimal coupling \cite{Chien:2021zle, Moroi:2000jr, Montefalcone:2022jfw, Freese:2014nla}). However, such an alteration in the theory, and hence, the dynamics, would be expected to leave its imprint on the duration of inflation and the process of reheating (including, but not limited to the generation of density perturbations). Of particular importance in this context is the equation of state parameter \cite{Zhou:2022ovp, Stein:2021uge}, and it remains to be seen if the aforementioned modifications are consistent with observations.

\section{An Observationally Consistent Model}
\label{sec:realistic_model}
We now venture to construct a scenario wherein the lacunae of the aforementioned toy models are trivially circumvented. Since the data constraints only the dynamics during the inflationary stage (or later), we begin with a discussion of this phase. In doing this, we are primarily guided by the need of naturally providing for sufficient slow-roll inflation and a graceful exit ({\em, i.e.}, without the necessity of invoking either super-Planckian scalar fields or undue fine-tuning). Simultaneously, the model must be in consonance with all observational data, \emph{i.e.,} with scalar and tensor perturbations. The Starobinsky model \cite{STAROBINSKY198099, STAROBINSKY1982175} is one such, and we begin this section with an examination of the same.

\subsection{Motivation for Starobinsky-like inflation}
\label{sec:3b-Starobinsky-inflation}
A modification of the Einstein-Hilbert Lagrangian, from $\sqrt{-g} R$ to $\sqrt{-g} f(R)$, where $f(R)$ is an arbitrary function, would be expected to significantly alter the dynamics, perhaps even to the extent of inducing a bounce. It should be noted, though, that a conformal transformation, taking one from this (Jordan) frame to another, can always be performed, thereby changing the form of the action. Indeed, as is well-known, as long as $f'(R) > 0$, the gravitational action can be recast (using a conformal transformation) as just the usual Einstein-Hilbert form but accompanied by a minimally coupled (to gravity) scalar field $\phi$ that serves to encode the information contained in the higher-derivative equation of motion that a generic $f(R)$ entails. The potential for the scalar field in this new frame (often termed as the Einstein frame) is, understandably, determined by the form of $f(R)$. For the case of
\beq f(R) = R + \frac{1} {6 m^2} R^2
\ , \label{fR}
\eeq
the ensuing potential has the form 
\begin{eqnarray}\label{eq:potential_Starobinsky_gen}
	V(\phi)=\frac{3}{4}m^2 \left(1-\exp{ \left( -\sqrt{\frac{2}{3}} \phi \right)}\right)^2.
\end{eqnarray}
Proposed by Starobinsky \cite{STAROBINSKY198099, STAROBINSKY1982175},
such a form of $f(R)$ in the Jordan frame or, equivalently, the
aforementioned $V(\phi)$ in the Einstein frame, has proved to be a
very successful theory of inflation. For large positive $\phi$, the
potential is indeed very flat and a slow-roll ensues. 

The Starobinsky form of $f(R)$ does not, of course, generate a bounce
on its own, as a violation of the NEC cannot be achieved. Rather, we
depend on the $\chi$-field to generate the bounce and, consider a form
of the potential inspired by that in
Eq. \eqref{eq:potential_Starobinsky_gen}, without ascribing it to a
$f(R)$. Indeed, as we argue below, concerns about stability too preclude a
purely geometric origin of the potential.
\subsection{Stability and the form of the potential}
It is well-known that inflationary expansion is, typically, a system
with enhanced dynamical stability, in that small fluctuations {\em en route} are of little importance, and the Universe proceeds towards an attractor solution.  Contraction, though, often exhibits the opposite behavior; as the Universe contracts, even the smallest fluctuations in the system can amplify relative to the background energy density, potentially leading to the disruption of the homogeneous background. One such example is the inherent anisotropic stress; generated quantum mechanically, it grows as $\propto a^{-6}$. During inflationary expansion, as the background energy density $3 H^2$ remains almost constant, anisotropic stress decays and does not cause any harm to the system. On the other hand, during inflationary contraction, the energy density of the anisotropic stress diverges
and tends to lead to the BKL instability \cite{Belinsky:1970ew, Karouby:2010wt, Karouby:2011wj, Bhattacharya:2013ut, Cai:2013vm, Ganguly:2021pke}. In our context, then, a symmetric
inflationary bouncing model, by construction, is liable to be
unstable.

Stabilization of the contracting phase could be achieved, for example, by incorporating ekpyrotic contraction as opposed to an inflationary one.  With the background energy density increasing faster than $a^{-6}$, the anisotropic stress can never dominate. This, by definition, implies a contracting phase differing from the expanding one and necessitates choosing a potential that is asymmetric about the point of bounce.

\subsection{The model}

With a Starobinsky-like potential for expansion and an
ekpyrotic-like for the contraction, all that remains is to be determined is the form that would facilitate the bounce. As we have seen earlier, a bounce can be achieved by having a $V_1(\phi)$ with a local maximum; we remain with the simple form considered earlier, namely a Gaussian form for $V_2(\chi)$ and $V_1(\phi)=A-B\phi^2$, with the values of $A$, $B$ depending on the location of the bounce and the value of the variable $n$ (that determines the evolution of $\rho_\chi$). In other words, we have

 \begin{eqnarray}\label{eq:pot_Starobinsky_inf_bounce_ekpy}
		&&V_1(\phi) =
		\left\{
		\barr{lclcl}
		\dis \alpha\ e^{ - \sqrt{\beta} \, \phi},\quad \beta \geq 6 &\qquad\quad& \phi \geq \phi_*, &\qquad\quad& (\text{Ekpyrotic contraction})
		\\[1.5ex]
		\dis   A_s - B_s(\phi-\phi_0)^2, & & \phi_* < \phi < \phi_i, && (\text{Bounce})
		\\[1.5ex]
		\dis \frac{3}{4}m^2 \left(1-  e^{-\sqrt{2/3} \, \phi}
		\right)^2, && \phi \leq \phi_i,  && (\text{Slow-roll expansion}),
		\earr
		\right.\\\nonumber
	&&	\text{and}\\
		&&V_2(\chi)=P_s \exp\left(-Q_s (\chi -\chi_0)^2 \right),
		\label{eq:pot_Starob_chi}
 \end{eqnarray}

where the constants $A_s$, $B_s$, $P_s$, $\phi_*$, $\alpha$, and $Q_s$
are determined in a manner similar to that used previously. For
detailed expressions, see \ref{appendix_S}.

\begin{figure}[H]
	\centering
	\includegraphics[height=0.41\textwidth,width=0.495\textwidth]{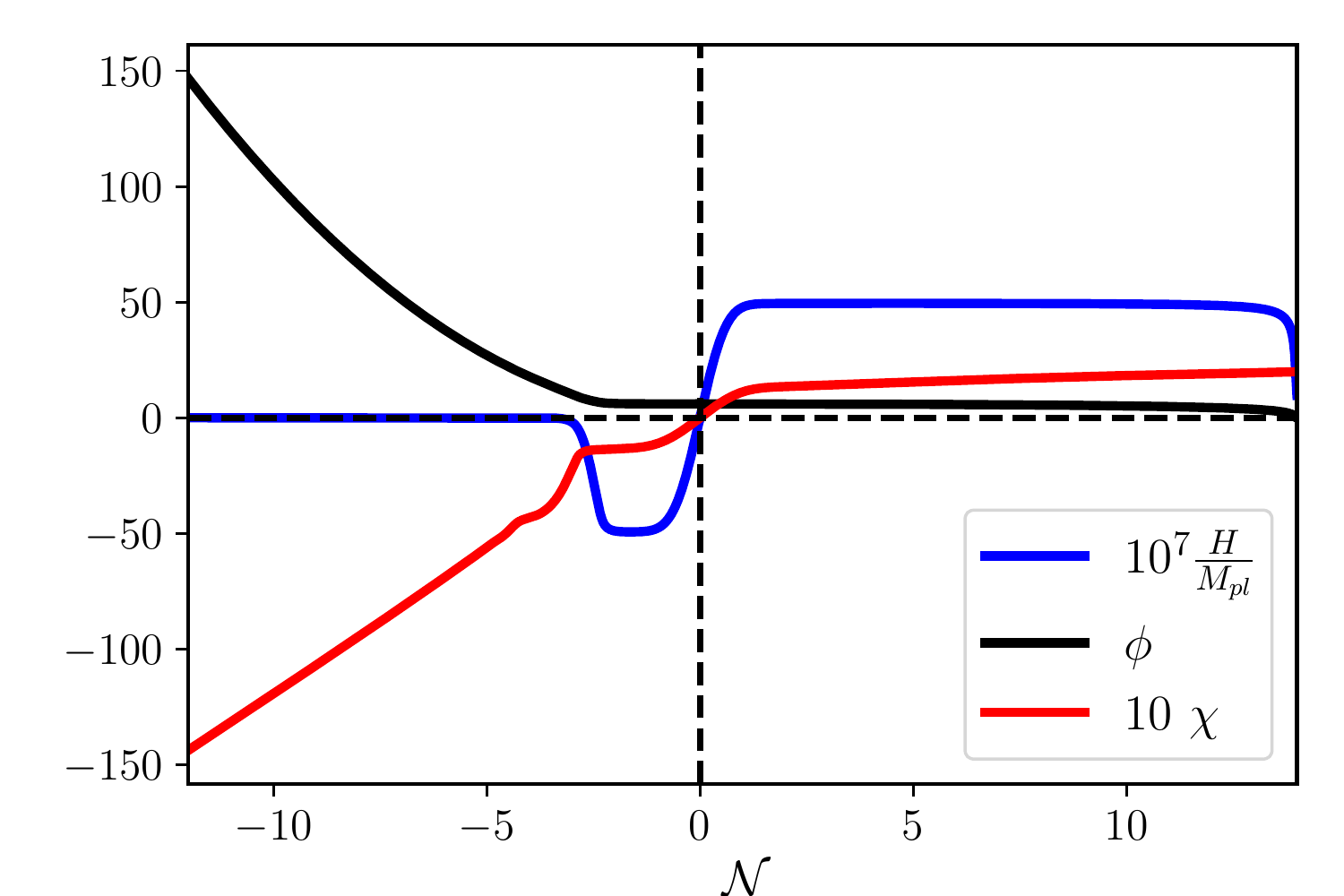}
	\includegraphics[height=0.4\textwidth,width=0.495\textwidth]{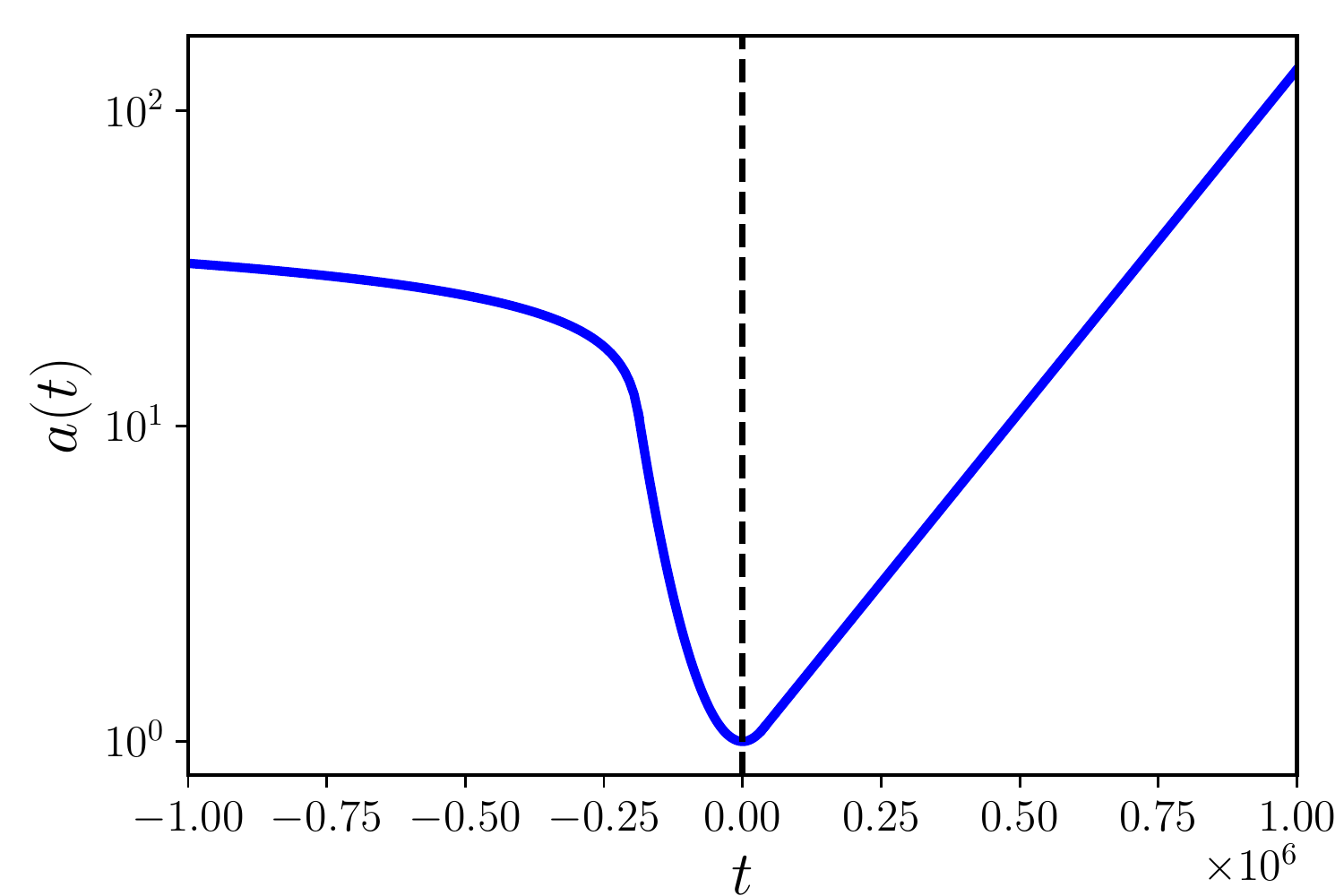}
	\caption{(Left) the evolution of the Hubble parameter $H$ and the fields $\phi$ and $\chi$ with $\mathcal{N}$ and, (Right) the evolution of scale factor $a$ with cosmic time $t$, obtained for potential described by
		Eqs. \eqref{eq:pot_Starobinsky_inf_bounce_ekpy} and \eqref{eq:pot_Starob_chi} for the parameters, $m=10^{-5}$, $n=7$, $\beta=6.5$ and $\phi_i=6$.}
	\label{fig:H_phi_chi_ekpy_staro}
\end{figure}

In Fig. \ref{fig:potential_staro_ekpy_inflation}, we show the form of
$V_1(\phi)$ for a particular choice of the parameters. As for the
dynamics, the asymptotic forms for $H(\mathcal{N})$ and $\phi(\mathcal{N})$ are easy to
determine starting with the equations of motion and the form of
potential in Eq. \eqref{eq:pot_Starobinsky_inf_bounce_ekpy}.  With a
Starobinsky-like inflation been well-studied in the literature, we
concentrate on the contracting phase. To begin with, let us impose the
initial conditions that, in the far past ($|\mathcal{N}|\gg1$ and $\mathcal{N}<0$), and solve the equations of motion given by Eqs. \eqref{eq:eqn1_in_e_N_fold} and \eqref{eq:eq_of_motion_phi_chi}.
To be specific, we have, for $|\mathcal{N}|\gg1$ and $\mathcal{N}<0$ (Ekpyrotic contraction phase)
\begin{eqnarray}{\label{asymp_negative}}
	H^2 &\approx& C_1  \exp{\left(\frac{-\beta \mathcal{N}^2}{2}\right)},\\
	\phi(\mathcal{N}) &\approx& \phi_e-\frac{1}{\sqrt{2\beta}}(\mathcal{N}^2-\mathcal{N}_e^2),
\end{eqnarray}
and, for ${\cal N} \gg \mathcal{N}_i$ (slow-roll expansion)
\begin{eqnarray}{\label{asymp_positive}}
	H^2 &\approx& \frac{m^2}{4} \left(1-\frac{3}{3 \xi_i -2(\mathcal{N}^2-\mathcal{N}_i^2)}\right)^2,\\
	\phi(\mathcal{N}) &\approx&
	\sqrt{\frac{3}{2}}\log{\left(-\frac{2}{3}(\mathcal{N}^2-\mathcal{N}_i^2)-\exp{(\sqrt{\frac{2}{3}}\phi_i)}\right)}.
\end{eqnarray}
\begin{figure}[t!]
	\centering    \includegraphics[height=0.5\textwidth,width=0.65\textwidth]{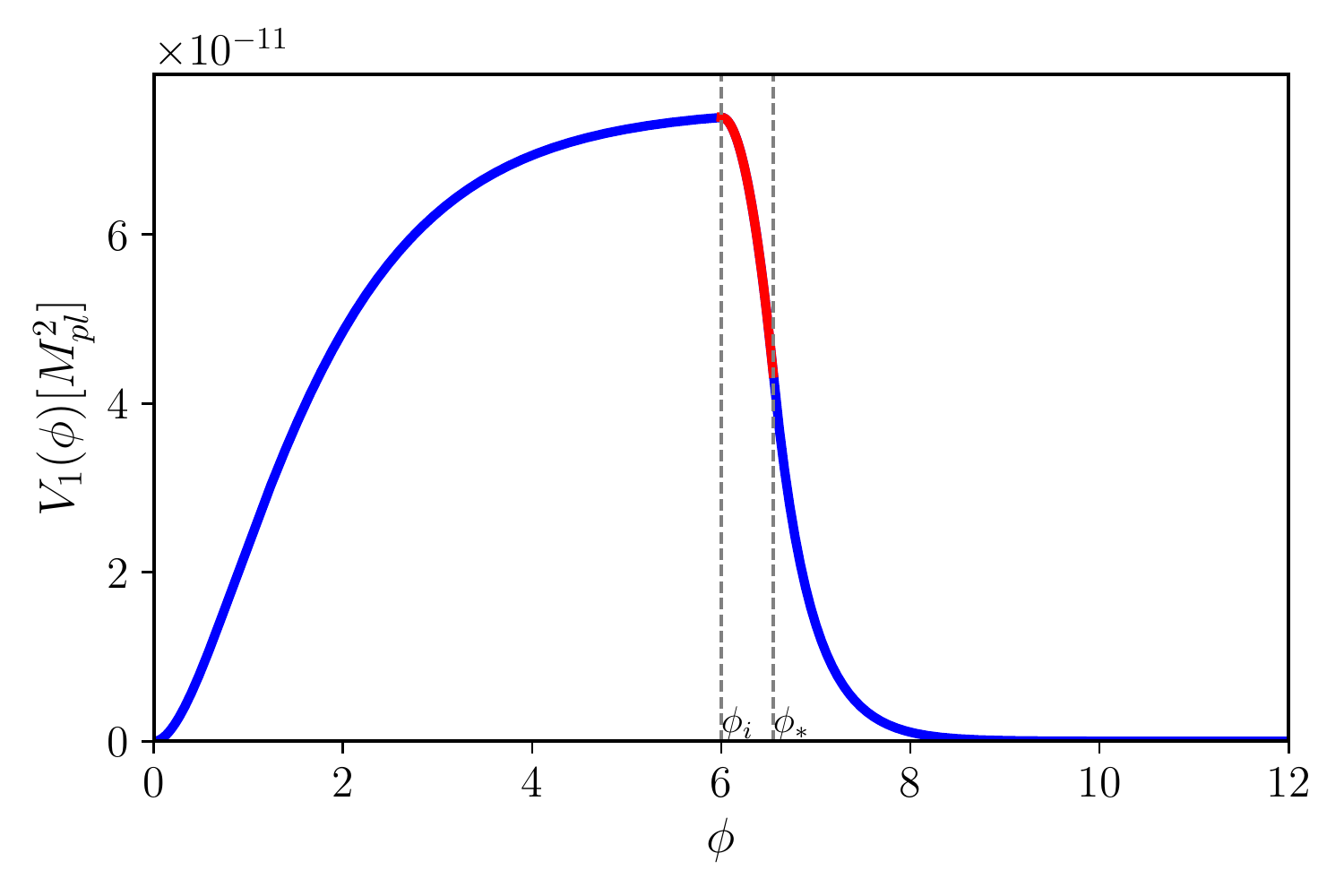}
	\caption{The potential $V_1(\phi)$ described in Eq. \eqref{eq:pot_Starobinsky_inf_bounce_ekpy} for the parameters, $m=10^{-5}$, $n=7$, $\beta=6.5$ and $\phi_i=6$. The red part in the inset corresponds to the bouncing phase, while the blue parts to the right (left) correspond to the contraction (expansion) phases, respectively. }
	\label{fig:potential_staro_ekpy_inflation}
\end{figure}
Here, $\phi_e = \phi({\cal N} = {\cal N}_e$) denotes the initial value of the
field $\phi$ (in the extreme
ekpyrotic contraction regime) while $\phi_i$ is the value
at the onset of slow-roll inflation at $\mathcal{N}=\mathcal{N}_i$ and $\xi_i = \exp{\left(\sqrt{\frac{2}{3}}\phi_i\right)}
$.
The constant $C_1$ is, in general, a
function of $\beta,\ \alpha,\ \phi_e$ and $\mathcal{N}_e$. 
The exact behavior can be obtained numerically and is depicted in
Fig. \ref{fig:H_phi_chi_ekpy_staro}.

In either of the two asymptotic phases, $V_2(\chi)$ is small and
$\chi$ contributes little to the evolution of $H$ and $\phi$. The
field $\chi$ itself develops as a free field, barring the influence of
the minimal coupling to gravity.

Close to the bounce, though, there is a qualitative change.  As
Eq. (\ref{asymp_positive}) suggests, the field $\phi({\cal N})$
drops to a relatively small value and, consequently, $V_1(\phi)$
increases. This immediately slows down the rate of change of $\phi$.
Almost simultaneously, the field $\chi$ too tends to stall (a consequence of
both its own potential $V_2(\chi)$ kicking in as well as the interplay with
$H({\cal N})$ and, through it, $\phi({\cal N})$). And, finally, thanks to the
wrong-sign kinetic term for the $\chi$, the bounce occurs.

It should be noted that such a construction will not result in qualitative changes in perturbations. We will in fact, demonstrate later that, in our instance, the initial condition for the scalar and tensor perturbations will be imposed during the bounce and not during the deep contraction.


\subsection{Observational signatures}
\label{pert}
One of the key issues that the inflationary paradigm has successfully addressed is the generation of small-scale perturbations as fluctuations (presumably of a quantum mechanical origin) in the inflaton field, which, subsequently, are amplified (as well as set up gravitational perturbations) leading to the generation of large-scale structure formation. It has been argued that models of bounce fail to address this with sufficient clarity owing to possible divergences at bounce.
\begin{figure}[H]
	\centering
	\includegraphics[height=0.55\textwidth,width=0.75\textwidth]{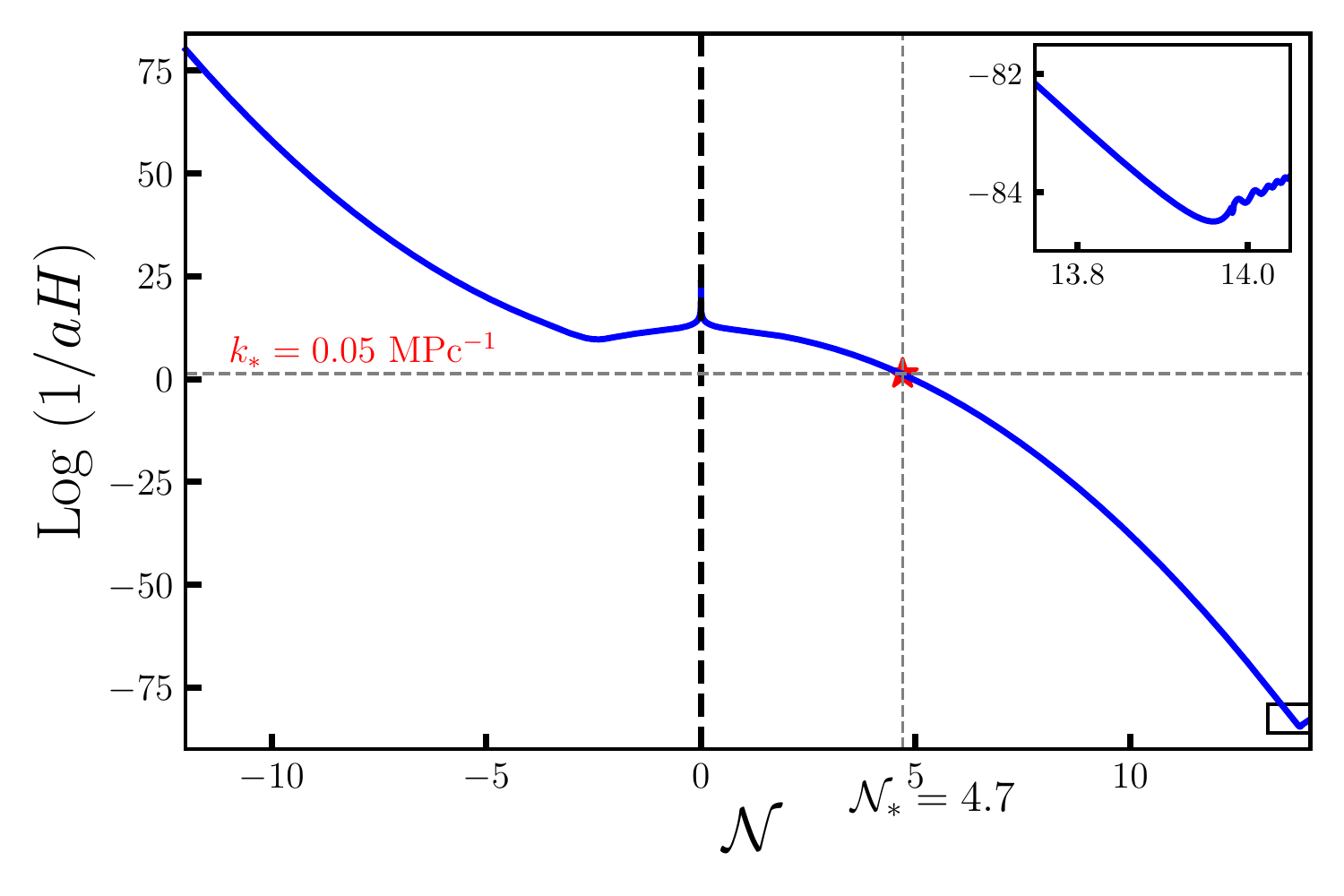} 
	\caption{ The evolution of comoving Hubble Horizon i.e., $\ln{\left({1}/{a H}\right)}$ with $\mathcal{N}$, obtained for the potentials described by Eqs. \eqref{eq:pot_Starobinsky_inf_bounce_ekpy} and \eqref{eq:pot_Starob_chi}, and parameters, $m=10^{-5}$, $n=7$, $\beta=6.5$ and $\phi_i=6$. Here red point corresponds to the pivot scale ($k_*=0.05\text{Mpc}^{-1}$), leaving the comoving Hubble horizon. In the subplot, the comoving Hubble horizon during reheating solution is magnified.}     \label{fig:hubblehorizon_Starobinsky_inflation_perturbation_1}
\end{figure}
It should be noted, though, that in such cosmologies, the behavior of perturbations can be highly dependent on the details of the bounce mechanism. For example, in certain ekpyrotic \cite{Raveendran:2018yyh, Levy:2015awa, Xue:2010ux, Wands:2008tv, Gordon:2002jw, Hwang:2002ks, Martin:2001ue} or matter bounce \cite{Agullo:2016tjh} models, perturbations do tend to diverge at the bounce; after the bounce, however, as the Universe enters the expansion phase, the perturbations can re-enter the linear regime and continue to evolve. As we argue below, an appropriate choice of the initial conditions would ensure that, in the scenario under discussion, the perturbations remain within the linear regime throughout the evolution without experiencing significant divergence. 

In studying the growth of perturbations, conventionally, the initial
conditions (Bunch-Davies) are imposed at an epoch where the modes of
interest are well inside the horizon. In the present context, the
bounce $(\mathcal{N}=0)$ is the unique epoch when the comoving Hubble
horizon diverges (thereby rendering all modes to be inside), and,
hence leads itself as a natural candidate for the imposition of the
initial conditions.  Additionally, for a scale leaving the Hubble
horizon at $\mathcal{N} = \mathcal{N}_*$, the mode remains within the
Hubble radius in the domain $-\mathcal{N}_* \leq \mathcal{N} \leq
\mathcal{N}_*$. During this time, as $k/(a H) \gg 1,$ and the
curvature and the entropic perturbations are nearly decoupled, the
approximate sub-Hubble solution for the curvature perturbation for the
comoving $k$-mode can be written as
\begin{eqnarray}
	&|\mathcal{R}_k| \simeq \frac{1}{\sqrt{2k} \,z(\mathcal{N})},\quad  z(\mathcal{N}) \equiv a(\mathcal{N}) \sqrt{2 \epsilon(\mathcal{N})},\quad |\mathcal{N}|<\mathcal{N}_*,&
\end{eqnarray} 
where $\epsilon(\mathcal{N}) \equiv - H'/{(\mathcal{N} H)}$ is the
slow-roll parameter. By design, the second field's influence is
limited to the curvature perturbation only around and immediately
following the horizon crossing. Choosing $\mathcal{N}_* >
\mathcal{N}_i$, with $\mathcal{N}_i$ being the onset of slow-roll
inflation, however, means that the Hubble crossing for a particular
mode lies within the $\phi$ field-dominated region, \emph{i.e.}, the
slow-roll inflationary region. As a result, in this case, the effect
of the $\chi$ field can, again, be disregarded, and the curvature perturbations behave similarly to those in
slow-roll inflation. When the horizon is crossed, \emph{i.e.}, $\mathcal{N} >
\mathcal{N}_*$, the modes enter the super-Hubble region, and, the
curvature perturbations freeze. Similar arguments are valid for the
tensor perturbation as well. In short, the very construction of the
theory guarantees that the curvature and tensor perturbation behave
similarly to those in slow-roll inflation. 
For $\mathcal{N}_* < \mathcal{N}_i$, the validity of this approximation is not guaranteed, causing the perturbations to behave differently than they would in slow-roll inflation

The pivot scale, $k_{\text{p}} = 0.05\,\text{Mpc}^{-1},$ crosses the
horizon $\sim 50$ e-folds prior to the end of inflation; this duration, in terms of e-N-folds, is equivalent to $\mathcal{N}_{\text{end}}^2 - \mathcal{N}_*^2 \sim
100$. This requirement of $\mathcal{N}_{\text{end}}>10$ often imposes
a condition on the value ($\phi_i$) of the field at the onset of the
inflation. For example, $\phi\sim 14.2$ at the pivot scale for chaotic inflation and $\sim 5.2$ for Starobinsky inflation, thereby imposing a lower bound on $\phi_i$ for either case. 

In conclusion, it is important to choose $\phi_i$ (equivalently
$\phi_0$) in such a way that the pivot scale leaves the Hubble horizon
very early in the inflationary phase. This ensures that the
perturbations grow analogously to those in usual slow-roll inflation,
thereby satisfying the observational constraints almost trivially. The
effect of bounce, however, might be observed at higher cosmological
scales (or, equivalently, for $k \ll k_p$), as these left the horizon
well before inflation could influence them significantly. While this
may even be confirmed from the study of the non-Gaussianity parameters,
this is beyond the scope of the current work, and we reserve this for a
future endeavor. Similar is the task of probing the contracting phase,
and thus the bounce itself, from observations. This could, presumably,
be done by imposing Bunch-Davies conditions before the bounce (while
ensuring that the observable modes still remain well inside the Hubble
horizon) and evolving the perturbations through the bounce. This, too, forms part of ongoing efforts.  

It should be noted that, due to the occurrence of a non-singular bounce at an exceedingly high energy scale, direct experimental testing is likely to be challenging. In order to obtain evidence, it is necessary to identify its observational consequences. This matter has been extensively discussed in the literature, and a plausible approach is to examine primordial perturbations \cite{Gasperini:1992em, Khoury:2001zk, Brustein:1994kn, Lyth:2001pf, Liu:2010fm, Xia:2014tda}
\section{Minimizing the ghost instability}\label{sec:4-removing-ghost}

The way we have formulated the analysis so far, the second term on the right-hand side of Eq. \eqref{Eq:genbounceH} implies a negative kinetic energy term. Since such a term normally leads to a ghost problem (at least as long as the fields are decoupled, as is the case here), this aspect needs a careful examination. While an absolute removal of the ghost is well-nigh impossible, we shall attempt a much less ambitious outcome, namely minimizing its effect on accessible physics. Consider, for example, an {\em ad-hoc} modification of Eq. \eqref{Eq:genbounceH} to
\begin{eqnarray}\label{eq:ghostH}
	H^2=f_1(\mathcal{N})+q(\mathcal{N})f_2(\mathcal{N}). 
\end{eqnarray}
Assuming it is possible to invoke, somehow, a term $q(\mathcal{N})$ such that it is negative {\em only} around the bounce and positive elsewhere, presumably, the ghostly behavior could be confined to the bounce epoch. Furthermore, since $f_2({\mathcal N})$ was seen to have a negligibly small impact away from the bounce, it stands to reason that this continues to be so unless $q({\mathcal N})$ grows away from the bounce. These twin requirements still leave a wide choice for the possible functional forms of $q({\mathcal N})$. For the sake of convenience (so that we may smoothly carry through the preceding quantitative analysis of the bounce mechanism), we normalize $q({\mathcal N}) = -1$ at the bounce and demand that, asymptotically away from it, $q({\mathcal N}) \to 1$.

While many choices are still possible, we choose to be economic and make do with one that is already present in theory. Remembering that diffeomorphism invariance of the underlying theory demands that $q(\mathcal{N})$ can be introduced only through a field, {\em viz.} $q(\phi(\mathcal{N}))$, we posit
\begin{eqnarray}\label{eq:coupling_function}
	q(\phi)=1-2\exp\left(-\frac{\alpha}{M_{\text{pl}}^2}(\phi-\phi_0)^2\right),
\end{eqnarray}
where $\alpha$ is a dimensionless constant. Since $q(\phi)$ is negative only for $|\phi - \phi_0| \leq \sqrt{\ln 2/\alpha} $, it might, naively, be argued that the period wherein the system faces a ghost instability may be reduced by choosing a larger $\alpha$.
While the introduction of the $q(\phi)$ might circumvent (at least in a limited sense) the problem of ghost, it would, of course, alter the dynamics (as this coupling will contribute significantly in the equations of motion). For all the good features of our model to be preserved, we need to ensure that such an introduction does not significantly affect the late time (\emph{i.e.,} away from the bounce) evolution of the scale factor. For example, starting with the action

	\begin{eqnarray}\label{eq:actiongen_removing_ghost}
		\mathcal{S}&=&\frac{\Mpl^2}{2}\int{\rm d}^4{\rm \bf x} \sqrt{-g} \left(R-\partial_\mu \phi \partial^\mu \phi -  2V_1(\phi)-q(\phi)\partial_\mu \chi \partial^\mu \chi-2V_2(\chi)\right),
	\end{eqnarray}

The corresponding equations of motion are
\begin{eqnarray}
	&& H^2=\frac{2\mathcal{N}^2(V_1+V_2)}{\left(6\mathcal{N}^2-\phi^{\prime 2}-q\chi^{\prime 2}\right)},\\
	&&\frac{H'}{H}=-\frac{\left(\phi^{\prime 2}+q\ \chi^{\prime 2}\right)}{2\mathcal{N}},\\
	&&\phi'' + \left( 3 \mathcal{N} + \frac{ H'}{H}-\frac{1}{\mathcal{N}}\right)\phi'+\frac{\mathcal{N}^2}{H^2}V_{1,\phi}-q_{,\phi}\frac{\chi^{\prime 2}}{2}=0,\\
	&&\chi'' + \left( 3 \mathcal{N} + \frac{ H'}{H}-\frac{1}{\mathcal{N}}\right)\chi'+ \frac{\mathcal{N}^2}{q H^2}V_{2,\chi}+\frac{q_{,\phi}}{q}\ \phi'\chi'=0.
\end{eqnarray}
As an illustrative example, consider the case of the Starobinsky model (with parameters identical to the previously studied case) but augmented by $q(\phi)$ of Eq. \eqref{eq:coupling_function}.

\begin{figure}[t!]
	\centering
	\includegraphics[height=0.5\textwidth,width=.65\textwidth]{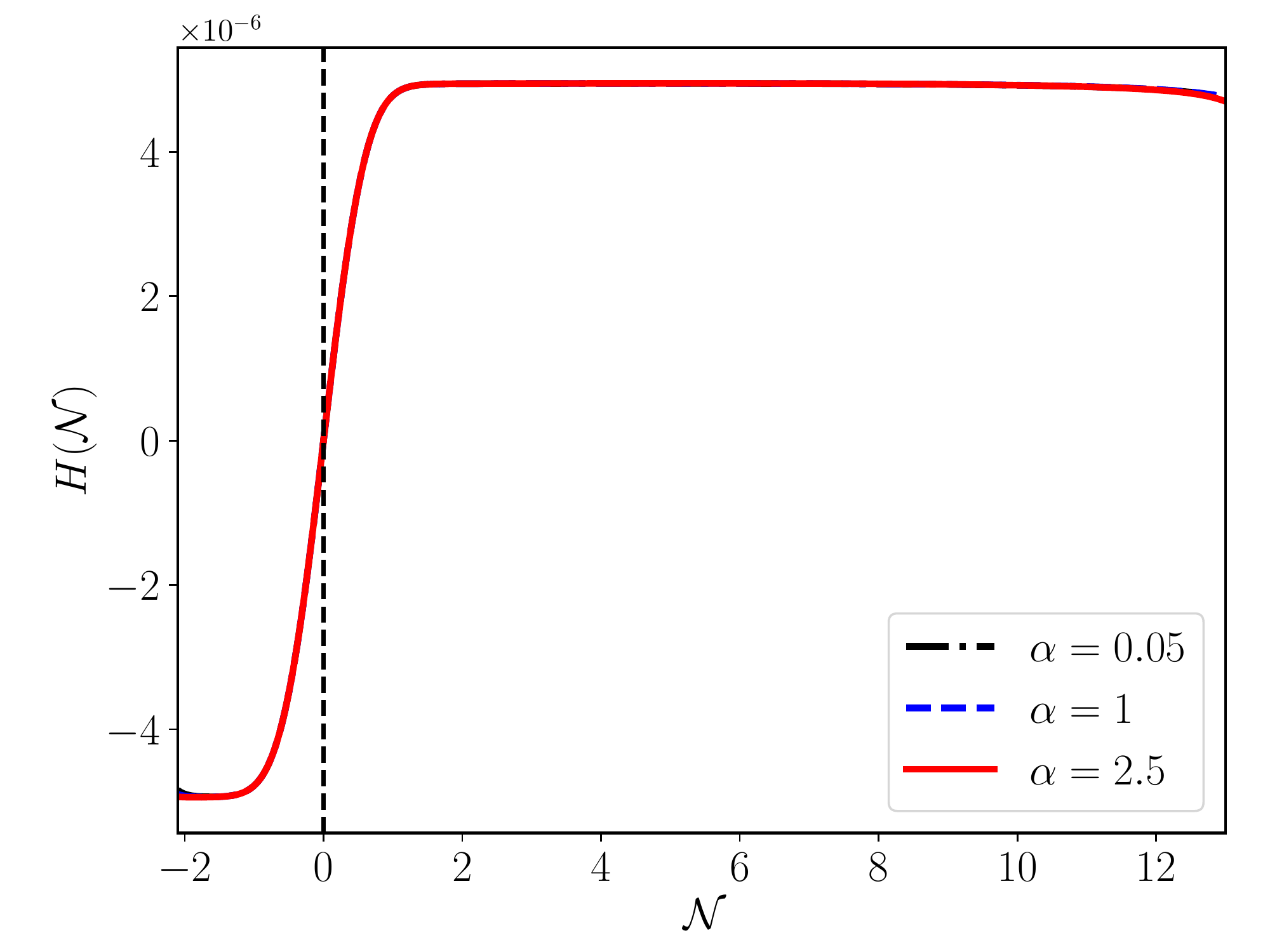}\\
 \includegraphics[height=0.3\textwidth,width=.45\textwidth]{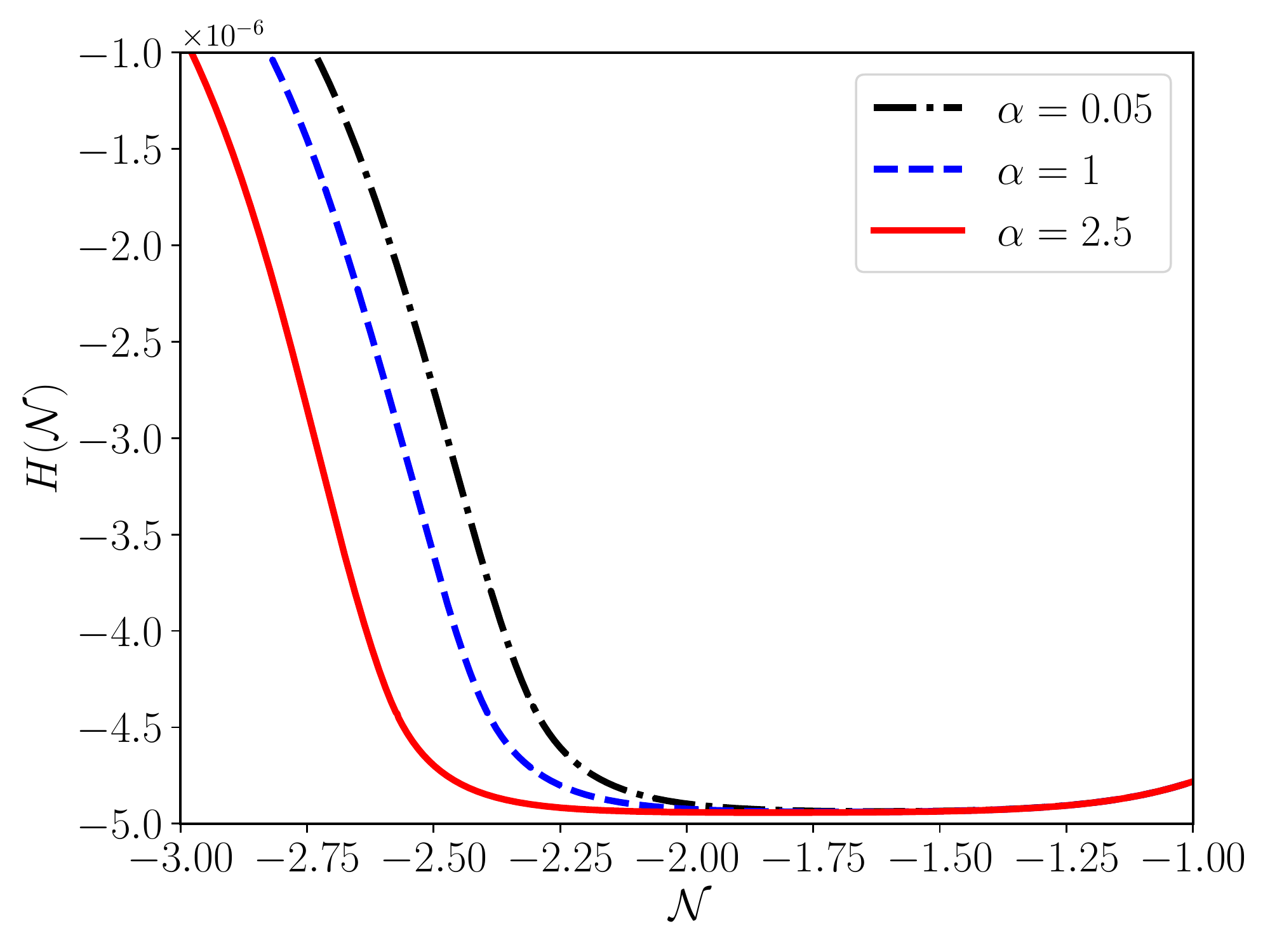}
	\includegraphics[height=0.3\textwidth,width=.45\textwidth]{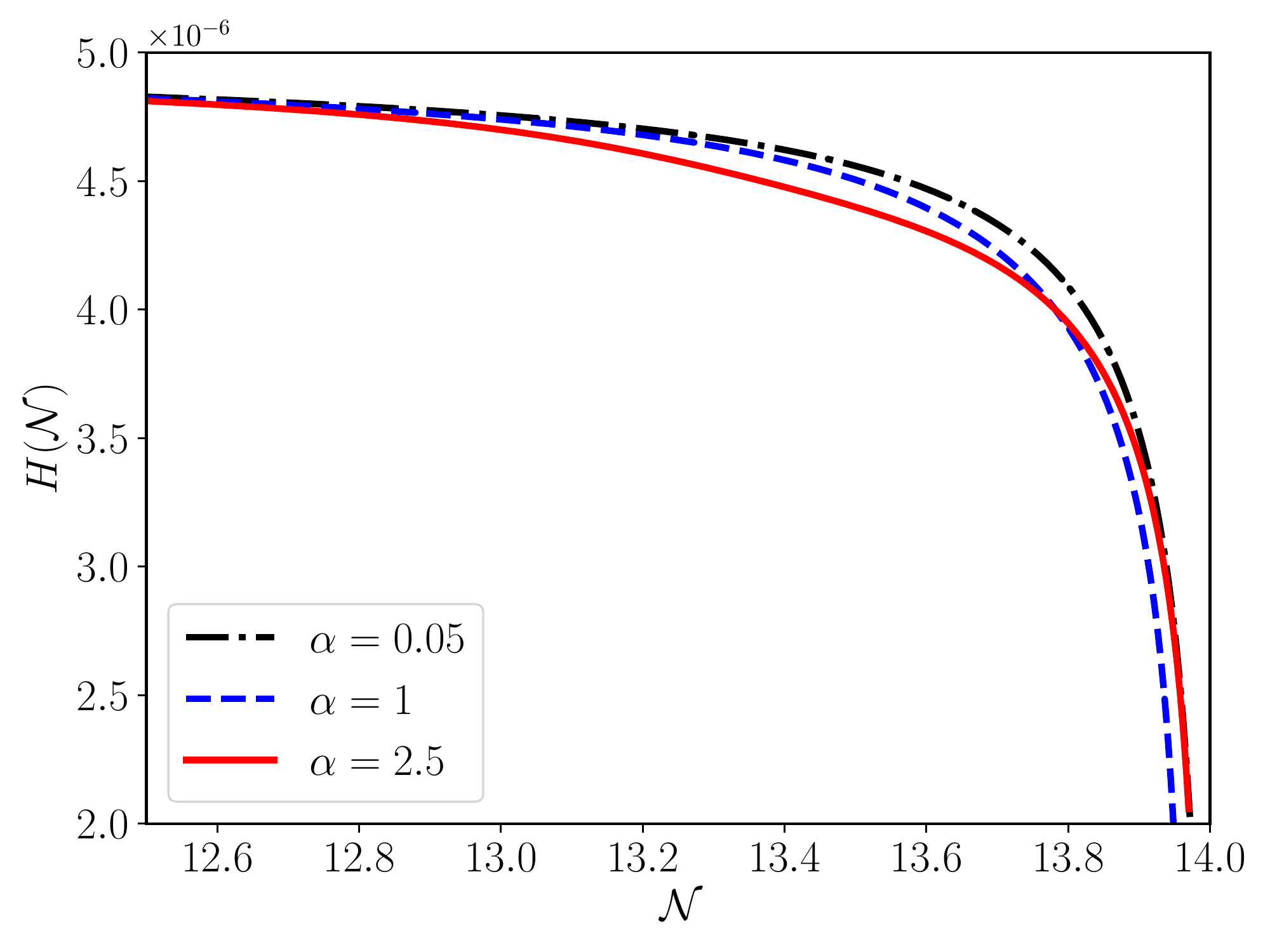}
	\caption{ (Top) The evolution of Hubble parameter $H(\mathcal{N})$ as a function of $\mathcal{N}$.  Zooming on to the evolution of $H$ during the contracting phase (bottom left) and the end of inflation (bottom right) are shown. For $\alpha\leq2.5$, the choice of coupling function introduced in Eq. \eqref{eq:coupling_function} can mitigate the ghost instability without much affecting the dynamics of the proposed model. }
	\label{fig:minimising_ghost}
\end{figure}

\noindent We numerically compute the evolution of the Universe, and the results are depicted in Fig. \ref{fig:minimising_ghost} for various choices of the parameter $\alpha$. For $\alpha=1$, the end of inflation occurs at $\mathcal{N}\simeq 10$; the pivot scale is required to leave the Hubble horizon at least 50 e-folds earlier, namely at $\mathcal{N}\simeq 5$. In this figure, it can be seen that, for $\alpha>1$, the transition happens $\mathcal{N}<10$. However, the dynamics of the Hubble parameter remains unaffected, with the only significant change arising in the contracting regime. It may be noted that for $\alpha>2.5$, the dynamics of the fields are greatly affected by the coupling term. 

\begin{figure}[h]
	\centering
	\includegraphics[height=0.45\textwidth,width=.65\textwidth]{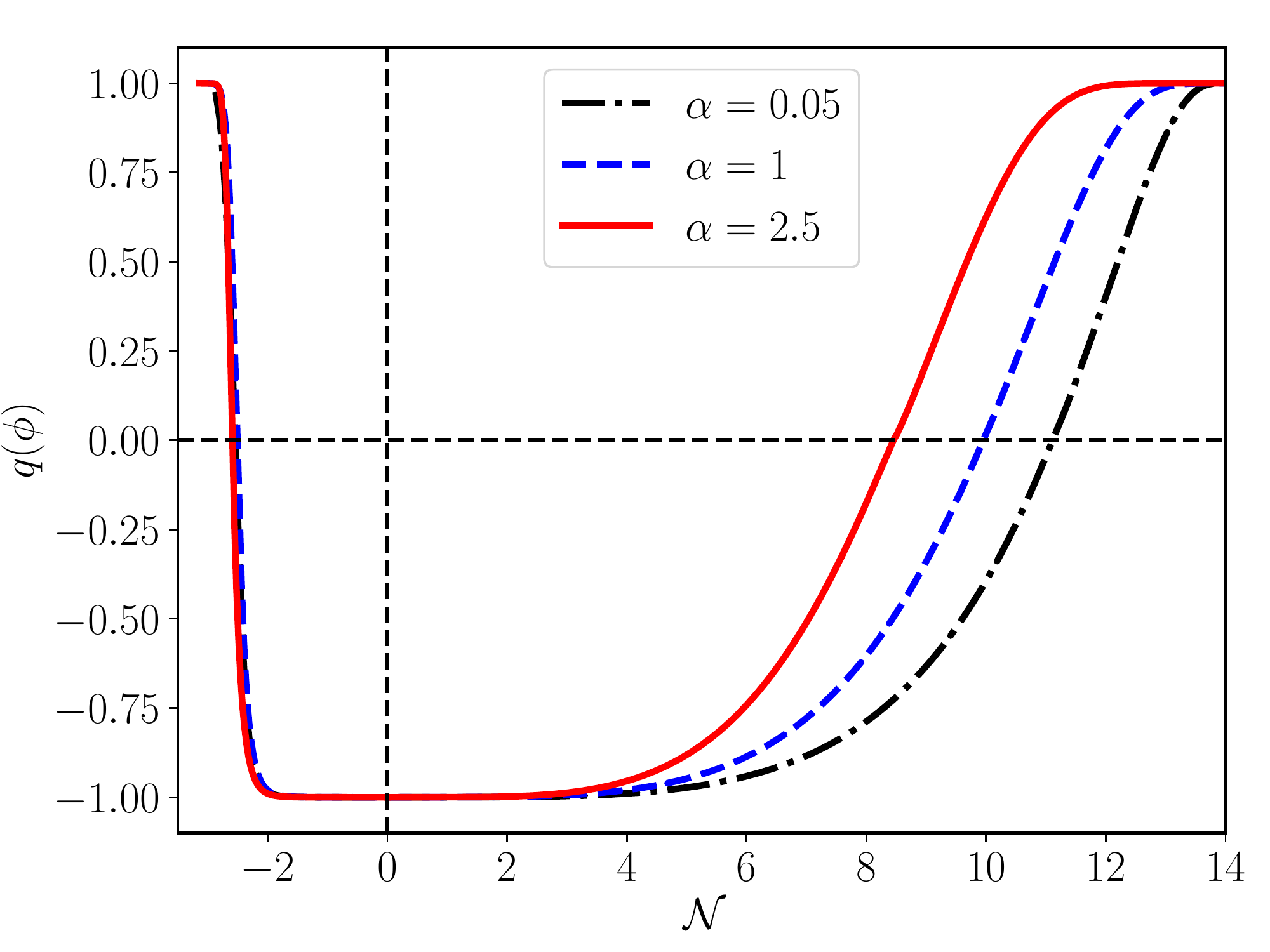}
	\caption{The evolution of coupling function $q(\phi)$ as a function of $\mathcal{N}$. Please note that for $\alpha>2.5$, the dynamics of the fields were greatly affected by the coupling term. For $\alpha\leq2.5$, the choice of coupling function introduced in Eq. \eqref{eq:coupling_function} can mitigate the ghost instability without much affecting the dynamics of the proposed model.}
	\label{fig:minimising_ghost_q}
\end{figure}

As can be seen in Fig. \ref{fig:minimising_ghost_q}, the transition of the function $q(\phi)$ (from negative to positive value) depends on the value of $\alpha$; the larger its value, the steeper is the transition. Hence, for $\alpha<2.5$, the choice of coupling function introduced in Eq. \eqref{eq:coupling_function} can mitigate the ghost instability without significantly affecting the dynamics of the proposed model.

These features are specific to the chosen inflationary model and cannot be generalized merely on the basis of the above analysis. It may be noted that the functional form of $q(\phi)$ assumed in Eq. \eqref{eq:coupling_function}, which can be thought to be a toy model, is what the analysis above hinges on. As a result, it is possible to compare and contrast various functional forms of $q(\phi)$ to get the best outcome, which we reserve for later efforts.

The above can, of course, be extended by generalizing the coupling function. Doing this would, largely, affect the dynamics close to the bounce epoch. Consequently, we postpone such a study for later.

\section{Summary and conclusions}\label{sec:6-conclusion}
The Classical bounce paradigm is one of the scenarios with the potential to resolve issues that plague standard inflationary cosmology. However, the construction of a bouncing model within the framework of standard gravity action necessitates the violation of the null energy condition, which often leads to ghost and gradient instabilities \cite{Kobayashi:2016xpl, Libanov:2016kfc, Ijjas:2016vtq, Kolevatov:2017voe, Banerjee:2018svi, Cai:2016thi, Cai:2017dyi, Mironov:2018oec, Easson:2011zy, Sawicki:2012pz}. Another major issue with a large class of bounce scenarios is that, in general, if the initial conditions are prescribed well before the bounce, the contracting phase is extremely sensitive to initial conditions. (This is true for all models with the exception of slowly contracting solutions such as 
\emph{ekpyrosis} \cite{Garfinkle:2008ei, Levy:2015awa}.) Consequently, anisotropic perturbations can grow fast, potentially breaking the overall homogeneity and isotropy, a phenomenon referred to as the Belinski-Khalatnikov-Lifshitz (BKL) instability \cite{Belinski:2009wj, Belinski:2014kba, belinski_henneaux_2017, Cai:2013vm, doi:10.1080/00018737000101171}. While such problems can, in principle, be circumvented, this typically involves going beyond canonical theories for gravity, invoking non-minimal couplings such as Horndeski theories~\cite{Banerjee:2018svi,  Cai:2016thi, Cai:2017dyi, Cai:2013vm, Dobre:2017pnt,  Gleyzes-et-al-2015, Horndeski:1974wa, Ijjas:2016vtq, Ilyas:2020qja,  Kobayashi:2016xpl, Kobayashi:2019hrl, Kolevatov:2017voe,  Libanov:2016kfc, Quintin:2015rta, Li:2016xjb, Mironov:2018oec,  Cai:2012va, Zhu:2021whu, Kumar:2021mgc, Pinto-Neto:2011gko,  Nandi:2020szp, Nandi:2023ooo, Nandi:2018ooh, Nandi:2019xag, Nandi:2019xlj, Nandi:2020sif, Nandi:2020szp, Nandi:2022twa, Kothari:2019yyw, Oikonomou:2022irx, Odintsov:2021urx, Banerjee:2022xft, Battista:2020lqv, Barca:2021qdn, Barca:2021wmj, Montani:2018uay} and/or considering the framework of  generalized Galileon cosmology with higher-derivative terms\cite{Qiu:2013eoa, Qiu:2011cy, Osipov:2013ssa}, with each such proposition is associated other drawbacks.

In this work, we have attempted to construct a history of the Universe that incorporates a contracting phase, followed by a bounce and, then, to a rapidly accelerating phase. Potential BKL instabilities are avoided (as can be seen in Sec. \ref{sec:3b-Starobinsky-inflation}) by choosing the contracting phase to be an ekpyrotic one. The model incorporates two scalar fields, each coupling minimally to gravity, with the latter, in turn, being described by the usual Einstein-Hilbert action. The violation of the null energy condition is effected through one of the scalar fields contributing negatively to the total energy density by virtue of an effective kinetic term that turns negative for a short while around the bounce. This leads to a situation where the Friedmann equation is given by
$$3H^2= \rho_\phi + q(\mathcal{N}) \, \rho_\chi,$$ 
where $\rho_{\phi, \chi}$ are the individual contributions of the two scalar fields $\phi, \, \chi$ to the total energy density in the absence of any direct coupling between them. If the function $q(\mathcal{N})$ were to be negative (if only during a certain epoch), it opens the possibility of the two contributions canceling each other, allowing $H$ to vanish, signaling a transition phase between a contracting ($H < 0$) and an expanding ($H > 0$) phase, namely a bounce. The above could be achieved, for example, by postulating that $\rho_\phi \sim \left( a_0/a\right)^m $ and $\rho_\chi \sim \left( a_0/a\right)^n$. For $n>m$, the said cancellation would be achieved at $a(\mathcal{N}) \sim a_0$ (with $a_0$ signaling the minimal size that the Universe may ever have had).  On the other hand, for $a \gg a_0$, $\rho_\chi$ would be subdominant and the effect of a non-canonical $q(\mathcal{N})$ submerged (for details see  \ref{appendix_A}). 

In this paper, we discussed a more interesting case, in which the bounce is followed by an inflationary era. This entails a choice for the potentials for $\phi, \chi$.  While many choices are permissible, initially, we self-restricted the enormous freedom by demanding that the dynamics be symmetric about the bounce. Tying it up with popular scenarios of inflation, such as chaotic inflation \cite{Linde:1983gd}, is possible and has been achieved here, albeit at the cost of introducing an upward convex phase in $V(\phi)$ (around the bounce) and sewing it up seamlessly with the usual chaotic phase\footnote{The seemingly ad-hoc nature of the aforementioned sewing-up of the potential can be avoided entirely if one were to consider {\em natural inflation}~\cite{PhysRevLett.65.3233} instead. With the inflaton $\phi$ now being identified with a pseudo-Nambu-Goto-Boson (such as an axion), the oscillatory nature of its potential is exactly what is needed!}. The requisite post-bounce slow-roll inflation is achieved without any fine-tuning beyond what is inherent to the respective base models. As is well-known, an inflationary phase, apart from setting up adequate density perturbations, also serves to suppress potential BKL instabilities. 
This is liable to be grossly violated in the contracting phase (endemic to bounce) if the latter is symmetric to the inflationary expansion. Thankfully, such instabilities can be prevented if the contraction were ekpyrotic in nature. With the consequent modification of the potential (to one asymmetric about the bounce), the resultant evolution (namely contraction followed by bounce and subsequent inflation) is rendered extremely stable.

It is worth pointing out that, for all three cases, we considered an identical form of potential for the $\chi$ field, namely a Gaussian, with only the parameters changing between the scenarios. 
For n=6, one can show that this potential vanishes identically and, yet, bounce occurs.

What is, however, more important is that a negative $q(\mathcal{N}),$ typically introduces ghost instabilities. We suggested a possible way out of this problem by introducing a nontrivial coupling between the two scalars by virtue of modifying the kinetic term for $\chi$ to $\beta(\phi) (\partial_\mu \chi) (\partial^\mu \chi)$. If $\beta(\phi)$ is a function such that it turns negative only close to the bounce and $\beta(\phi)$ is positive away from it, one could confine the instability to only a narrow epoch in the history of the Universe. What is heartening is that $\beta(\phi)$ need not be fine-tuned.  Indeed, as we showed, a simple function of the inflaton potential $V(\phi)$ is all we need. Of course, the introduction of such a coupling between the fields should not alter the dynamics of the proposed model. However, the change in the evolution is only quantitative and not a qualitative one, as is demonstrated by explicit computations. 
This coupling function can, of course, be generalized, but we reserve a detailed analysis of the consequences for a future investigation. To summarise, we achieved the following 
\begin{itemize}
	\item By using a class of classical bouncing models with two scalar fields in the minimal Einstein configuration and by selecting a suitable asymmetric bouncing model, we were able to avoid the initial condition problem of the early Universe, i.e., our model is stable during both contracting and expanding phases, thereby avoiding crucial BKL instabilities. 
	\item In the minimal Einstein frame, the extant literature \cite{Raveendran:2018yyh, Levy:2015awa, Xue:2010ux, Wands:2008tv, Gordon:2002jw, Hwang:2002ks, Martin:2001ue} rarely discusses the divergences of scalar and tensor perturbations at the bounce. By setting the Bunch-Davies initial condition for the perturbations at the bounce, we ensure that our model is free from such divergences. In this configuration, the detailed effects of perturbations on both scalars and tensors remain to be investigated; we reserve this for future study. 
	\item We mitigate the ghost instability by confining the issue to the bounce by introducing a coupling function between the fields.
\end{itemize}
\section*{Acknowledgement}
MK is supported by a DST-INSPIRE Fellowship under the reference
number: IF170808, DST India. DN is supported by the DST-INSPIRE faculty fellowship program with faculty registration no.: IFA20-PH-255. DC and TRS acknowledge their respective grants (Reg No./IoE/2021/12/FRP) awarded by the IoE, University of Delhi.
\appendix
\appendix
\section{Simple bouncing models}\label{appendix_A}

Concentrating on the toy model mentioned in the introduction, {\em viz.}, one where the Hubble parameter takes the form
\begin{eqnarray}{\label{eq:hubblematter}}
	3H^2=\left( \frac{a_0}{a}\right)^m-\left( \frac{a_0}{a}\right)^n, \quad n>m.
\end{eqnarray}
with $a_0$ being the minimum value of scale factor and $n$ and $m$ being constants\footnote{In Refs. \cite{Raveendran:2017vfx, Raveendran:2018why}, the Hubble parameter has been expressed in terms of the conformal time $(\eta)$, whereas in this article we have expressed the same in terms of the scale factor.}, for $a\gg a_0$, for $a\gg a_0$, the second term can be neglected altogether, leading to 
\begin{eqnarray}{\label{eq:scalefactor_powerlaw}}
	a(t)\propto |t|^{ 2/m},
\end{eqnarray}
with the two branches being applicable of expanding ($t>0$) and contracting ($t<0$) phases, respectively. As the $\chi$ field is subdominant in this regime, it is easy to determine the potential for $\phi$ field that would lead to such a $a(t)$, namely 
\begin{eqnarray}\label{eq:pot_matter}
	V_1(\phi)\propto \exp{(-\sqrt{m}|\phi|)},
\end{eqnarray}
Around $a \sim a_0$, i.e., around the bounce, the second term of Eq. \eqref{eq:hubblematter} can no longer be neglected, and we have
\begin{eqnarray}
	\label{eq:phiprime_V_phifield}
	&&\phi'=\pm\sqrt{m}\mathcal{N} \left(1-e^{\frac{(-n+m) \mathcal{N}^2}{2}}\right)^{-1/2},
	\\
	&&V_1(\mathcal{N})=\left(1-\frac{m}{6}\right) e^{\frac{-m\mathcal{N}^2}{2}},\\
	\label{eq:chiprime_V_chifield}
	&&\chi'=\pm\sqrt{n}\mathcal{N} \left(-1+e^{\frac{(n-m) \mathcal{N}^2}{2}}\right)^{-1/2},\\&&V_2(\mathcal{N})=\left(\frac{n}{6}-1\right) e^{\frac{-n\mathcal{N}^2}{2}}.
\end{eqnarray}

\noindent Here, $V_1(\mathcal{N})\equiv V_1(\phi(\mathcal{N}))$ and $V_2(\mathcal{N})\equiv V_2(\chi(\mathcal{N}))$. Integrating the above differential equations, we have
\begin{eqnarray}\label{eq:powerlaw_phi}
	\phi &=&\phi_0\pm \frac{ 2 \sqrt{m} } { (n-m) } \arctanh{\left(\sqrt{1-e^{\frac{(m-n)}{2} \mathcal{N}^2}}\right)},\\
	\label{eq:powerlaw_chi}\chi &=& \chi_0\pm\frac{2\sqrt{n}}{(n-m)}\, \, \arctan{\left(\sqrt{e^{\frac{(n-m)}{2} \mathcal{N}^2}-1}\right)},
\end{eqnarray}
where $\phi_0$ and $\chi_0$ are the respective values of the fields at the bounce epoch ($\mathcal{N}$=0). The corresponding potentials are
\begin{eqnarray}
	\label{eq:pot_phi_powerlaw}
	V_1(\phi)&=&\left( \frac{6-m}{6}\right) \left(\cosh{\left( \frac{(n-m)}{2\sqrt{m}}(\phi-\phi_0)\right)}\right)^{\frac{2m}{m-n}},\nonumber\\&&\\
	\label{eq:pot_chi_powerlaw}   
	V_2(\chi)&=&\left( \frac{n-6}{6}\right) \, \, \, \left(\cos{\left( \frac{(n-m)}{2\sqrt{n}}(\chi-\chi_0)\right)}\right)^{\frac{2n}{n-m}}.\nonumber\\
\end{eqnarray}
It is reassuring that the expression for $V_1(\phi)$ smoothly interpolates between the two branches in Eq. \eqref{eq:pot_matter}. Furthermore, with
\begin{eqnarray}
	-\frac{\pi}{2}\leq \frac{(n-m)}{2\sqrt{n}}(\chi-\chi_0) \leq \frac{\pi}{2},
\end{eqnarray}
for $|\mathcal{N}|\gg 1$, the potential $V_2(\chi)$ can indeed be ignored compared to $V_1(\phi),$  ensuring that our formulation is self-consistent.
\section{Inflationary evolution}
\label{appendix_D}
Inflation is a brief period of accelerated expansion, and it provides a satisfactory resolution to the shortcomings of the hot big-bang model. The simplest yet the most successful model of inflation can be described by a single scalar field slowly rolling down the potential --- referred to as slow-roll inflation. The action corresponding to this scenario can easily be achieved by using a canonical scalar field with potential $V(\phi)$ minimally coupled to gravity as
\begin{eqnarray}\label{eq:general_inflation_action}
	\mathcal{S}=\frac{1}{2}\int{\rm d}^4{\rm \bf x} \sqrt{-g} \left(R- \partial_\mu \phi \partial^\mu \phi - 2V(\phi)\right),
\end{eqnarray}
and the corresponding equations of motion in the flat FLRW background can be written as 
\begin{eqnarray}
	3 H^2 = \frac{1}{2} \dot{\phi}^2 + V(\phi),\qquad \quad \ddot{\phi}+3H\dot{\phi}+V_{,\phi}=0,
\end{eqnarray}
with the slow-roll parameters defined as 
\begin{eqnarray}
	\epsilon \equiv -\frac{\dot{H}}{H^2},\qquad  \quad  \eta\equiv\frac{\dot{\epsilon}}{H\epsilon}.
\end{eqnarray}
As the field slowly rolls downs the potential for a sufficient time, one can immediately make approximations as  $(\epsilon \ll 1$ and $\eta \ll 1)$, known as the slow-roll approximations and the corresponding background equations reduce to
\begin{eqnarray}\label{eq:equation_with_slowroll}
	3H^2 \simeq V(\phi),\qquad \quad 3H\dot{\phi}\simeq - V_{,\phi}.
\end{eqnarray}
The above form can easily be written in terms of $\mathcal{N}$ as
\begin{eqnarray}\label{eq:inflationphiderivative}
	\frac{1}{\mathcal{N}}\frac{{\rm d} \phi}{{\rm d} \mathcal{N}} \simeq -\frac{V_{\phi}}{V}.
\end{eqnarray}
\noindent One can solve the above equation to find the time-dependence of the field and obtain the solution for the Hubble parameter as $3H^2 \simeq V(\phi).$ 
\subsection{Chaotic Inflation}\label{appendix_B}

In the case of a canonical field, minimally coupled to gravity with the chaotic potential
\begin{eqnarray}
	V(\phi)=\frac{1}{2}m^2\phi^2,
\end{eqnarray}
under slow-roll approximation, the Hubble parameter (using Eq. \eqref{eq:equation_with_slowroll}) can be written as,
\begin{eqnarray}\label{eq:equation_with_slowroll_chaotic}
	3H^2\simeq \frac{1}{2}m^2\phi^2
	=\frac{1}{2}m^2(-2\mathcal{N}^2+2\mathcal{N}_i^2+\phi_i^2),
\end{eqnarray}
where we choose the initial condition: $\phi(\mathcal{N}_i)\equiv\phi_i$.
\subsection{Starobinsky Inflation}
In the case of Starobinsky inflation, the potential in Einstein's frame is given by
\begin{eqnarray}
	V(\phi)=\frac{3}{4}m^2 \left(1-\exp{ \left( -\sqrt{\frac{2}{3}} \phi \right)}\right)^2,
\end{eqnarray}
under slow-roll approximation, Hubble parameter using Eq. \eqref{eq:equation_with_slowroll} can be written as,
\begin{eqnarray}\label{eq:equation_with_slowroll_starobinsky}
	H^2\simeq \frac{1}{4}m^2 \left(1-\frac{3}{\left(3\exp{\left(\sqrt{\frac{2}{3}}\phi_i\right)}-2(\mathcal{N}^2-\mathcal{N}_i^2)\right)}\right)^2,\\
\end{eqnarray}
where we choose the initial condition: $\phi(\mathcal{N}_i)\equiv\phi_i$.
It may be noted that in the case of conventional chaotic inflation Hubble parameter diverges as we go back in time, whereas as in the case of Starobinsky inflation, the Hubble parameter becomes constant $(\sim m/2)$.
\section{Bounce with Starobinsky Inflation}\label{appendix_S}

As stated earlier, the coefficients $A_s$ and $B_s$ appearing in Eq. \eqref{eq:pot_Starobinsky_inf_bounce_ekpy} are to determined by imposing continuity of the potential $V_1(\phi)$ at $\phi = \pm |\phi_0 - \phi_i|$. Defining,
\beq
\xi_i \equiv \exp{\left(\sqrt{\frac{2}{3}}\phi_i\right)},
\eeq
an explicit computation gives
\beq
\barr{rcl}
A_s & = & \dis \frac{m^2}{144 \xi_i^2} \left(144 \mathcal{N}_i^2 +108 \xi_i^2
-216 \xi_i +60\right),\\[3ex]
B_s&=& \dis  \frac{m^2}{144 \xi_i^2}
\left(28 n \mathcal{N}_i^2 +27 n \xi_i^2 -42 n \xi_i -3 (n+24)\right),\\
\earr
\eeq
The coefficients $P_s$, $Q_s$ and $\alpha$, on the other hand, are determined to be

\begin{eqnarray}
	P_s &=&  \dis \frac{m^2(n-6)\left(-3+3 \xi_i+2\mathcal{N}_i^2\right)^2}
	{8\left(3 \xi_i+2\mathcal{N}_i^2\right)^2},\\
	Q_s &=&  \dis
	\frac{24-n\left(3 \xi_i +2\mathcal{N}_i^2\right)
		\left(-3+3 \xi_i+2\mathcal{N}_i^2\right)}
	{4\left(-3+3\xi_i+2\mathcal{N}_i^2\right)
		\left(3 \xi_i+2\mathcal{N}_i^2\right)},\\
	\alpha &=& \exp{\left(\sqrt{\beta}\phi_*\right)}\left(A_s- B_s (\phi_0-\phi_*)^2\right),
\end{eqnarray}
where   

	\begin{eqnarray}
		\label{eq:parameter_Starobinsky2}
		\mathcal{N}_i^2& = & \dis \frac{1}{168 n} \left(
		\Xi 
		+n \left(126 \xi_i -81 \xi_i^2+ 37\right)+216
		\right)\\
		\phi_0 &= &
		\frac{\sqrt{6}\, \Xi +216\sqrt{6}+n \left(-252 \phi_i-81 \sqrt{6} \xi_i^2 + 126 \xi_i \left(2 \phi_i+\sqrt{6}\right)-19\sqrt{6}\right)}{252\  n \left(\xi_i-1\right)},\\
		\phi_* &=& \frac{-\sqrt{\beta} B_s + \sqrt{\beta B_s(\beta A_s + B_s)}+ \beta B_s \phi_0}{\beta B_s},\\
		\Xi^2 & \equiv & \dis n^2 \left(9 \xi_i \left(9 \xi_i-14\right)+19\right)^2
		+432 n \left(9 \xi_i \left(5 \xi_i-14\right)+107\right)+ 46656.
	\end{eqnarray}


\bibliographystyle{apsrev4-2}

\end{document}